\documentclass[twocolumn]{article}
\usepackage{graphicx} % Required for inserting images
\usepackage{amsmath}
\usepackage{amssymb}
\usepackage{amsfonts}
\usepackage{hyperref}
\usepackage{braket}
\usepackage{datetime}
\usepackage{xcolor}
\usepackage{bm}
\usepackage{tikz}
\usepackage{enumitem}
\usetikzlibrary{quantikz2}
\usepackage[bibstyle=nature,
            citestyle=numeric-comp,
            sorting=none,
            maxbibnames=5, minbibnames=1,
            doi=false,
            eprint=false,
            isbn=false,
            url=true]{biblatex}
\DeclareSourcemap{
  \maps[datatype=bibtex]{
    \map[overwrite]{
      \step[fieldsource=date, match=\regexp{\A(\d{4}).*\Z}, replace=\regexp{$1}]
      \step[fieldset=month, null]
      \step[fieldset=urldate, null]
      \step[fieldset=language, null]
      \step[fieldset=note, null]
      \step[fieldset=abstract, null]
      \step[fieldset=issue, null]
    }
    \map{
      \step[fieldsource=journal, match=\regexp{\AAnnals\sof\sOperations\sResearch\Z}, replace={Ann. Oper. Res.}]
      \step[fieldsource=journaltitle, match=\regexp{\AAnnals\sof\sOperations\sResearch\Z}, replace={Ann. Oper. Res.}]
      \step[fieldsource=journal, match=\regexp{\AIEICE\sTransactions\son\sFundamentals\sof\sElectronics,\sCommunications\sand\sComputer\sSciences\Z}, replace={IEICE Trans. Fundam. Electron. Commun. Comput. Sci.}]
      \step[fieldsource=journaltitle, match=\regexp{\AIEICE\sTransactions\son\sFundamentals\sof\sElectronics,\sCommunications\sand\sComputer\sSciences\Z}, replace={IEICE Trans. Fundam. Electron. Commun. Comput. Sci.}]
      \step[fieldsource=journal, match=\regexp{\APhysical\sReview\sLetters\Z}, replace={Phys. Rev. Lett.}]
      \step[fieldsource=journaltitle, match=\regexp{\APhysical\sReview\sLetters\Z}, replace={Phys. Rev. Lett.}]
      \step[fieldsource=journal, match=\regexp{\AFrontiers\sin\sQuantum\sScience\sand\sTechnology\Z}, replace={Front. Quantum Sci. Technol.}]
      \step[fieldsource=journaltitle, match=\regexp{\AFrontiers\sin\sQuantum\sScience\sand\sTechnology\Z}, replace={Front. Quantum Sci. Technol.}]
      \step[fieldsource=journal, match=\regexp{\Anpj\sQuantum\sInformation\Z}, replace={npj Quantum Inf.}]
      \step[fieldsource=journaltitle, match=\regexp{\Anpj\sQuantum\sInformation\Z}, replace={npj Quantum Inf.}]
      \step[fieldsource=journal, match=\regexp{\ANature\sCommunications\Z}, replace={Nat. Commun.}]
      \step[fieldsource=journaltitle, match=\regexp{\ANature\sCommunications\Z}, replace={Nat. Commun.}]
      \step[fieldsource=journal, match=\regexp{\AScience\sAdvances\Z}, replace={Sci. Adv.}]
      \step[fieldsource=journaltitle, match=\regexp{\AScience\sAdvances\Z}, replace={Sci. Adv.}]
      \step[fieldsource=journal, match=\regexp{\AMathematical\sProgramming\sComputation\Z}, replace={Math. Program. Comput.}]
      \step[fieldsource=journaltitle, match=\regexp{\AMathematical\sProgramming\sComputation\Z}, replace={Math. Program. Comput.}]
      \step[fieldsource=journal, match=\regexp{\AAdvanced\sQuantum\sTechnologies\Z}, replace={Adv. Quantum Technol.}]
      \step[fieldsource=journaltitle, match=\regexp{\AAdvanced\sQuantum\sTechnologies\Z}, replace={Adv. Quantum Technol.}]
      \step[fieldsource=journal, match=\regexp{\ANature\sReviews\sPhysics\Z}, replace={Nat. Rev. Phys.}]
      \step[fieldsource=journaltitle, match=\regexp{\ANature\sReviews\sPhysics\Z}, replace={Nat. Rev. Phys.}]
      \step[fieldsource=journal, match=\regexp{\AORSA\sJournal\son\sComputing\Z}, replace={ORSA J. Comput.}]
      \step[fieldsource=journaltitle, match=\regexp{\AORSA\sJournal\son\sComputing\Z}, replace={ORSA J. Comput.}]
      \step[fieldsource=journal, match=\regexp{\AJournal\sof\sthe\sPhysical\sSociety\sof\sJapan\Z}, replace={J. Phys. Soc. Jpn.}]
      \step[fieldsource=journaltitle, match=\regexp{\AJournal\sof\sthe\sPhysical\sSociety\sof\sJapan\Z}, replace={J. Phys. Soc. Jpn.}]
      \step[fieldsource=journal, match=\regexp{\ANew\sJournal\sof\sPhysics\Z}, replace={New J. Phys.}]
      \step[fieldsource=journaltitle, match=\regexp{\ANew\sJournal\sof\sPhysics\Z}, replace={New J. Phys.}]
      \step[fieldsource=journal, match=\regexp{\ANational\sScience\sReview\Z}, replace={Natl Sci. Rev.}]
      \step[fieldsource=journaltitle, match=\regexp{\ANational\sScience\sReview\Z}, replace={Natl Sci. Rev.}]
      \step[fieldsource=journal, match=\regexp{\ACommunications\sPhysics\Z}, replace={Commun. Phys.}]
      \step[fieldsource=journaltitle, match=\regexp{\ACommunications\sPhysics\Z}, replace={Commun. Phys.}]
      \step[fieldsource=journal, match=\regexp{\APhysical\sReview\sApplied\Z}, replace={Phys. Rev. Appl.}]
      \step[fieldsource=journaltitle, match=\regexp{\APhysical\sReview\sApplied\Z}, replace={Phys. Rev. Appl.}]
      \step[fieldsource=journal, match=\regexp{\AScientific\sReports\Z}, replace={Sci. Rep.}]
      \step[fieldsource=journaltitle, match=\regexp{\AScientific\sReports\Z}, replace={Sci. Rep.}]
      \step[fieldsource=journal, match=\regexp{\AIEEE\sTransactions\son\sComputer-Aided\sDesign\sof\sIntegrated\sCircuits\sand\sSystems\Z}, replace={IEEE Trans. Comput.-Aided Des. Integr. Circuits Syst.}]
      \step[fieldsource=journaltitle, match=\regexp{\AIEEE\sTransactions\son\sComputer-Aided\sDesign\sof\sIntegrated\sCircuits\sand\sSystems\Z}, replace={IEEE Trans. Comput.-Aided Des. Integr. Circuits Syst.}]
      \step[fieldsource=journal, match=\regexp{\AACM\sTransactions\son\sQuantum\sComputing\Z}, replace={ACM Trans. Quantum Comput.}]
      \step[fieldsource=journaltitle, match=\regexp{\AACM\sTransactions\son\sQuantum\sComputing\Z}, replace={ACM Trans. Quantum Comput.}]
      \step[fieldsource=journal, match=\regexp{\AQuantum\sInformation\sProcessing\Z}, replace={Quantum Inf. Process.}]
      \step[fieldsource=journaltitle, match=\regexp{\AQuantum\sInformation\sProcessing\Z}, replace={Quantum Inf. Process.}]
      \step[fieldsource=journal, match=\regexp{\APhysical\sReview\sResearch\Z}, replace={Phys. Rev. Res.}]
      \step[fieldsource=journaltitle, match=\regexp{\APhysical\sReview\sResearch\Z}, replace={Phys. Rev. Res.}]
      \step[fieldsource=journal, match=\regexp{\APhysical\sReview\sA\Z}, replace={Phys. Rev. A}]
      \step[fieldsource=journaltitle, match=\regexp{\APhysical\sReview\sA\Z}, replace={Phys. Rev. A}]
      \step[fieldsource=journal, match=\regexp{\AComputational\sOptimization\sand\sApplications\Z}, replace={Comput. Optim. Appl.}]
      \step[fieldsource=journaltitle, match=\regexp{\AComputational\sOptimization\sand\sApplications\Z}, replace={Comput. Optim. Appl.}]
      \step[fieldsource=journal, match=\regexp{\AQuantum\sMachine\sIntelligence\Z}, replace={Quantum Mach. Intell.}]
      \step[fieldsource=journaltitle, match=\regexp{\AQuantum\sMachine\sIntelligence\Z}, replace={Quantum Mach. Intell.}]
      \step[fieldsource=journal, match=\regexp{\AIEEE\sTransactions\son\sComputers\Z}, replace={IEEE Trans. Comput.}]
      \step[fieldsource=journaltitle, match=\regexp{\AIEEE\sTransactions\son\sComputers\Z}, replace={IEEE Trans. Comput.}]
      \step[fieldsource=journal, match=\regexp{\ASIAM\sJournal\son\sOptimization\Z}, replace={SIAM J. Optim.}]
      \step[fieldsource=journaltitle, match=\regexp{\ASIAM\sJournal\son\sOptimization\Z}, replace={SIAM J. Optim.}]
    }
    \map{
      \pertype{article}\pertype{inproceedings}\pertype{incollection}\pertype{book}\pertype{techreport}
      \step[fieldset=url, null]
    }
  }
}
\DeclareFieldFormat{nat:arxiv}{Preprint at \url{https://arxiv.org/abs/#1}}
\DeclareFieldFormat{nat:url}{\url{#1}}
\DeclareFieldFormat[article,inproceedings,incollection,report,techreport,misc,online]{title}{#1}
\DeclareFieldFormat{titlecase}{\ifcurrentfield{title}{\MakeSentenceCase*{#1}}{#1}}
\DeclareBibliographyDriver{misc}{%
  \usebibmacro{bibindex}%
  \usebibmacro{begentry}%
  \usebibmacro{author/editor+others/translator+others}%
  \setunit{\labelnamepunct}\newblock
  \usebibmacro{title}%
  \newunit\newblock
  \iffieldundef{eprint}
    {\printfield[nat:url]{url}}
    {\printfield[nat:arxiv]{eprint}}%
  \setunit{\addspace}%
  \iffieldundef{year}{}{\printtext[parens]{\printfield{year}}}%
  \usebibmacro{finentry}}
\DeclareBibliographyAlias{online}{misc}
\DeclareFieldFormat{labelnumberwidth}{\mkbibbrackets{#1}}
\usepackage{geometry}
\usepackage{caption}
\usepackage{titlesec}
\usepackage{float}
\usetikzlibrary{shapes.geometric, arrows.meta, positioning, calc}

\usepackage{booktabs}
\usepackage{multirow}
\usepackage{array}
\usepackage{algorithm}
\usepackage{algpseudocode}
\usepackage{siunitx} % optional, but gives cleaner formatting

\newsavebox{\solidbox}
\newsavebox{\dashedbox}
\sbox{\solidbox}{\tikz[baseline=-0.5ex]{\draw[line width=0.9pt] (0,0) -- (0.5,0);}}
\sbox{\dashedbox}{\tikz[baseline=-0.5ex]{\draw[line width=0.9pt, dashed] (0,0) -- (0.5,0);}}

\DeclareMathOperator*{\argmin}{argmin}

\renewcommand{\thesection}{\Roman{section}}
\renewcommand{\thesubsection}{\Alph{subsection}}
\renewcommand{\thesubsubsection}{\arabic{subsubsection}}
\makeatletter
\renewcommand{\p@subsection}{\thesection.}
\renewcommand{\p@subsubsection}{\thesection.\thesubsection.}
\makeatother

\titleformat{\section}
  {\centering\normalfont\bfseries\MakeUppercase}
  {\thesection.}
  {1em}
  {}

\titleformat{\subsection}
  {\centering\normalfont\small\bfseries}
  {\thesubsection.}
  {1em}
  {}

\titleformat{\subsubsection}
  {\centering\normalfont\small\itshape}
  {\thesubsubsection.}
  {1em}
  {}

\usepackage[normalem]{ulem}

\DeclareRobustCommand{\claudefix}{\tikz[remember picture,overlay,baseline=0pt]{%
  \coordinate (cfx) at (0,0.7ex);%
  \path let \p1=(cfx), \p2=(current page.center) in
    \pgfextra{\ifdim\x1<\x2 \gdef\cfxside{west}\gdef\cfxshift{12mm}\else \gdef\cfxside{east}\gdef\cfxshift{-12mm}\fi};%
  \node[text=orange, inner sep=0pt, xshift=\cfxshift] at (cfx -| current page.\cfxside) {$\bigstar$};}}

\title{\textbf{Quantum optimization beyond qubit count by low-energy subspace sampling}}
\author{
  Mateo Alonso \textsuperscript{1, 2, *},
  Gonzalo Blázquez-Gil\textsuperscript{1, \dag},
  Pablo Díez-Valle \textsuperscript{1, \ddag}\\ [6pt]
  \textit{\small \textsuperscript{1}\, Instituto Tecnológico de Galicia, Cantón Grande 9, Planta 3, 15003 A Coruña, Spain}\\
  \textit{\small \textsuperscript{2}\,Universidade da Coruña, Computer Architecture Group, CITIC, 15071 A Coruña, Spain} \\
  \small \textsuperscript{*}\,\texttt{malonso@itg.es}, \textsuperscript{\dag}\,\texttt{gblazquez@itg.es}, \textsuperscript{\ddag}\,\texttt{pvalle@itg.es}
}

\date{}

\begin{document}

\twocolumn[
  \begin{@twocolumnfalse}
    \maketitle
    \begin{abstract}

    On current digital quantum devices, standard one-qubit-per-variable encodings restrict quantum optimization algorithms to problem sizes where classical solvers typically excel. To avoid this constraint, we introduce Quantum Subspace Sampling Descent (QSSD), an iterative heuristic that decouples the number of problem variables from qubit count, while preserving knowledge of the problem’s global structure. QSSD drives the energy descent using a fixed-angle Quantum Approximate Optimization Algorithm as a low-energy sampler. As the angles are precomputed, each iteration requires just a single circuit execution. We test QSSD on IBM Heron and Nighthawk quantum devices with up to 152 qubits on instances of higher-order unconstrained binary optimization and Max-cut problems. On the largest instance with 10000 variables, QSSD reaches an approximation ratio above 0.95 with 100 circuit executions. We provide both exact and heuristic classical baselines, namely greedy, branch-and-bound solver, and simulated annealing. While our approach still does not match the performance of the best classical baseline, it shows promise in a slower runtime scaling with problem size given a fixed budget. This suggests a regime of large problems where subspace sampling on quantum devices could become competitive, as we also demonstrate how improvements in hardware will lead to better algorithm performance. QSSD is thus a practical heuristic for combinatorial optimization problems that helps bring quantum optimization a step closer to utility.

    \end{abstract}
    \bigskip
  \end{@twocolumnfalse}
]
%%%%%%%%%%%%%%%%%%%%%%%%%%%

\section{Introduction}

Combinatorial optimization is the task of finding the assignment of a finite set of discrete variables $\mathbf{z} \in \{-1,+1\}^{m}$ that minimizes a given cost function $E(\mathbf{z})$. Problems of this form are ubiquitous across science, engineering and industry, including, for instance, portfolio selection, vehicle routing, job scheduling, etc.~\cite{papadimitriou1982combinatorial}. The number of candidate configurations grows exponentially with the number of variables $m$, making an exhaustive search impractical even for modest sizes.

Classical approaches to combinatorial optimization fall into two broad families. Exact solvers, typically based on branch-and-bound and cutting-plane techniques, such as HiGHS~\cite{HiGHs},  Gurobi~\cite{gurobi} or CPLEX~\cite{cplex}, guarantee optimality, but their practical performance depends critically on the structure and sparsity of the instance, with exponential runtime in worst-case scenarios. Heuristics, in contrast, trade guarantees for scalability, returning approximate solutions within practical runtimes. Well-known heuristic optimization methods are simulated annealing~\cite{simulated_annealing}, tabu search~\cite{glover_tabu_1989} or parallel tempering~\cite{swendsen_replica_1986, hukushima_exchange_1996}. The quality of the optimal solution they achieve is not uniform with approximation ratios that depend strongly on the problem instance, and degrades on rugged, densely coupled or higher-order landscapes where the search stalls in local minima. 

Quantum optimization is an alternative optimization paradigm that takes as its starting point the mapping of the cost function onto a diagonal spin Hamiltonian $H_{C}$, in which each spin variable $z$ is replaced by a Pauli-$Z$ operator so that $H_{C}\lvert \mathbf{z}\rangle = E(\mathbf{z})\lvert \mathbf{z}\rangle$~\cite{Abbas_2024}. The optimal assignment of variables then corresponds to the ground state of the resulting Ising or higher-order model. In the case of Higher-Order Unconstrained Binary Optimization (HUBO) problems, they are mapped to 
\begin{equation}
\label{eq:hubo}
    H_{C} = \sum_i^{m} h_i Z_i + \sum_{i > j}^{m} J_{ij} Z_iZ_j + \sum_{i>j>k}^{m} K_{ijk} Z_i Z_j Z_{k}\,,
\end{equation}
where $Z_i$ corresponds to the Pauli-$Z$ matrix acting on the $i$-th qubit, and  $h_i, J_{ij}, K_{ijk} \in \mathbb{R}$ are coefficients defined by the instance. Typical approaches implement nonlinear terms as multi-qubit phase rotations, and assign one qubit per problem variable. With current digital quantum processors offering on the order of $10^{2}$ qubits and coherence times limiting circuit depth, the variable-qubit correspondence is a critical limitation, and hardware demonstrations of quantum optimization algorithms remain confined to a few hundred variables~\cite{alevras_mrna_2024, nodar2025scalingvariationalquantumeigensolver, Romero2025, Zhu_2026, Sachdeva_2026}. This number of variables is quite short of the instance sizes routinely addressed in many practical applications and handled by the classical methods described above. Before quantum optimization can compete on solution quality, it must therefore first reach comparable problem sizes. Two families of strategies have been proposed to close this gap. 

The first compresses the encoding itself, so that a single qubit carries information about many variables. Qubit-efficient schemes exploit the exponential dimension of Hilbert space by associating problem configurations with the basis states or amplitudes of a register using up to $\mathcal{O}(\log m)$ qubits~\cite{tan2021qubitefficient}, an idea subsequently refined and extended to broader classes of instances and protocols~\cite{advqt2024_2300309, pra2024_109_052441}. A complementary route is the Pauli Correlation Encoding (PCE), which achieves a polynomial compression by assigning each problem variable to a distinct Pauli string, so that the $m$ variables are read out from correlators of order $k$ measured on a qubit register scaling only as $m^{1/k}$~\cite{sciorilli_towards_2025}. This approach has been demonstrated on instances of up to $10^4$ variables on 14 qubits on trapped-ion hardware, although the PCE hardware results are comparable to greedy post-process from random initial strings at that size~\cite{yoshioka_pce_2026}, and applied to problems such as portfolio optimization~\cite{soloviev_portfolio_2025} and mRNA structure prediction on IBM devices~\cite{friedhoff_pce_2026}. However, these proposals have scalability issues due to, among other reasons, the cost of amplitude estimation and variational optimization. These issues limit the conducted experiments with these new encoding schemes to fewer than 30 qubits, which is a regime that can be emulated classically in short time.  

The second family decomposes the problem instead, solving subinstances of tractable size on the quantum device and reassembling their solutions. Partitioning schemes, for example, cut the interaction graph into weakly coupled blocks, optimize each block on the processor and stitch the partial assignments together~\cite{li_dcqaoa_2023, zhou_qaoa2_2023, ushijima_multilevel_2021, ponce_decomposition_2025}. One of the main shortcomings of partitioning is that the broken couplings are treated approximately or ignored, which is untenable in high-density cases. Another class of decomposition methods is sub-QUBO decomposition, which repeatedly optimizes a subset of variables while the rest are fixed to a reference solution. These subsets can be chosen based on different criteria, such as single-flip impact~\cite{booth_qbsolv_2017, rosenberg_iterative_2016}, ease of embedding~\cite{okada_embedding_2019}, the coupling structure of the problem~\cite{bass_decomposition_2021, zhao_clustering_2025}, the disagreement across a pool of solutions~\cite{rosenberg_iterative_2016, bass_decomposition_2021, atobe_subqubo_2022}, or greedily or at random within a tabu search~\cite{moussa_tabu_2021}. To solve the resulting subproblems, there are approaches using a quantum annealer~\cite{booth_qbsolv_2017, okada_embedding_2019, bass_decomposition_2021, atobe_subqubo_2022}, classical tabu search~\cite{booth_qbsolv_2017, rosenberg_iterative_2016, atobe_subqubo_2022} or variational QAOA~\cite{moussa_tabu_2021, zhao_clustering_2025}.

\begin{figure*}[t]
\centering
\includegraphics[width=1\linewidth]{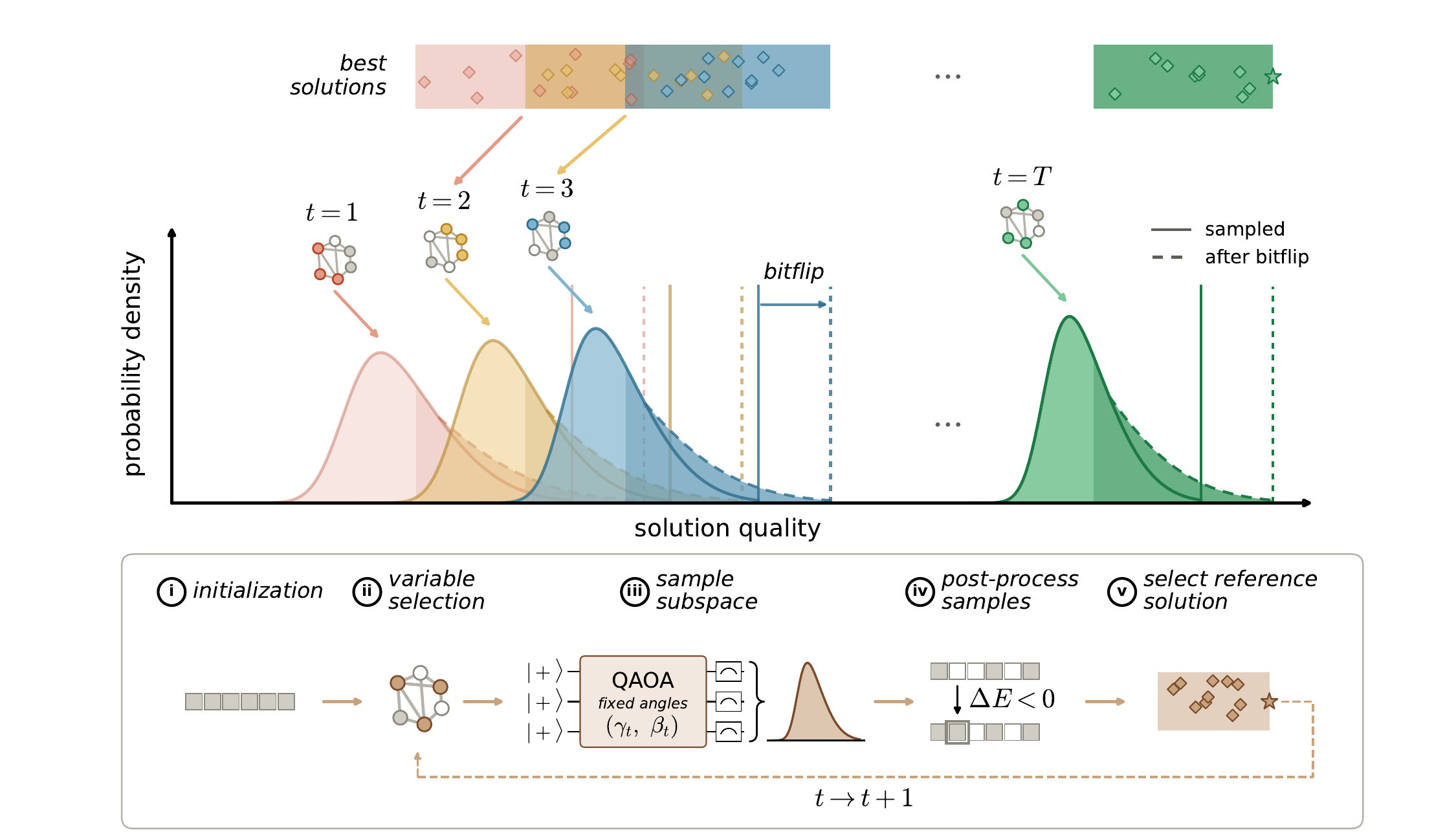}
\caption{Schematic of the QSSD algorithm. Top: evolution of the distribution of the solution quality of the samples obtained at different iterations $t=1,2,3,\dots,T$, each for a different subset of free variables (represented by the colored nodes in the graphs) while keeping the rest fixed to the previous solution value (white and gray nodes). Solid and dashed vertical lines mark the best sample before and after the bitflip post-processing, respectively. The shaded tail of each distribution corresponds to the samples that enter the population of best configurations represented in the top section. These populations are updated at every iteration and determine the variables selected for the following iteration of the algorithm. The final solution (star) and output of QSSD is taken as the best configuration in the population in the final iteration. Bottom: steps of one iteration of QSSD, (i) initialization of the reference solution, (ii) selection of $n$ variables from the scores computed on the population of best solutions, (iii) sampling of the reduced subproblem with a QAOA circuit with analytically computed angles $(\boldsymbol{\gamma}_t,\boldsymbol{\beta}_t)$, (iv) bitflip post-processing of the lowest-energy samples, which keeps the changes that result in lower energy values, and (v) update of the population and of the reference solution, which is used in the next iteration, $t+1$.}
\label{fig:algorithm_schematic}
\end{figure*}

Here we introduce Quantum Subspace Sampling Descent (QSSD), a heuristic that combines iterative subspace selection with the low-energy state sampling perspective on the Quantum Approximate Optimization Algorithm (QAOA). At each iteration, a subset of $n \ll m$ variables is selected on the basis of a relevance score computed from the population of low-energy configurations found so far, while the remaining variables are frozen to their values in the current solution, yielding a reduced HUBO on $n$ spins that is mapped to an $n$-qubit QAOA circuit. A schematic of the QSSD algorithm is included in Fig.~\ref{fig:algorithm_schematic}. Note that this is not a standard partition method, as it cuts no couplings and the frozen variables are considered in the reduced subproblem exactly, and more closely resembles the sub-QUBO decomposition methods. Also, in contrast with recursive elimination schemes~\cite{bravyi_rqaoa_2020, finzgar_qiro_2024}, QSSD does not permanently freeze variable values, any variable can be selected again in a future iteration, and does not require as many high-quality qubits as there are variables in the problem. 
Crucially, the circuit is not optimized variationally, as its angles are fixed analytically from the coefficient statistics of the reduced instance following Refs.~\cite{Sureshbabu2024parametersettingin, Farhi2022quantumapproximate}, so that a single circuit per iteration suffices and trainability issues such as the barren-plateau problem are bypassed entirely. The measurements performed on the QAOA circuit provide low-energy samples that are used to update the best solutions, establishing a monotonically descending upper bound on the global optimum. The quantum device thus acts as a low-energy sampler rather than a subproblem solver, and the full distribution it returns serves a dual purpose: escaping local minima and informing which variables to optimize next. This is unlike the sub-QUBO decomposition methods reviewed~\cite{booth_qbsolv_2017, rosenberg_iterative_2016,okada_embedding_2019,bass_decomposition_2021, zhao_clustering_2025,atobe_subqubo_2022, moussa_tabu_2021}, which only keep a single bitstring per subproblem.

We show that this design allows problems far larger than the available qubit count to be addressed with good quality on present-day hardware. Using up to 152 qubits of superconducting quantum devices provided by IBM, QSSD reaches an approximation ratio higher than $0.95$ on a 10000-variable HUBO instance with 100 quantum circuit evaluations, outperforms the quantum algorithm PCE on Gset instances, and is able to get competitive results with a standard classical exact solver, HiGHS. In addition, while the simulated annealing heuristic reaches better solutions than QSSD within the range of maximum iterations (50-100) tested, it shows worse runtime scaling with problem size. Furthermore, ablation studies isolate the contribution of the quantum sampler, which is shown to beat random sampling and an exact solver applied to the same subspaces when sufficient layers are used, confirming that the value of the quantum step lies in its low-energy distribution rather than in the optimum alone. In line with this, solution quality further improves with the number of qubits and depth of the quantum circuits employed, so that advances in device size and fidelity are expected to translate into better solutions.

The remainder of the paper is organized as follows. Section~\ref{sec:results} describes the QSSD algorithm in detail, motivates and evaluates the performance of QAOA with precomputed angles as a low-energy sampler and benchmarks the protocol on IBM quantum hardware against other classical and quantum optimizers. Section~\ref{sec:discussion} discusses these results, the utility regimes of QSSD and its limitations. Section~\ref{sec:methods} defines the problem classes employed in tests, the QAOA circuit construction tailored to heavy-hex HUBO problems and the classical and quantum solvers used in the comparisons.

\section{Results}
\label{sec:results}

\subsection{Quantum Subspace Sampling Descent}

The main contribution of the work is the introduction of the Quantum Subspace Sampling Descent algorithm (QSSD), a heuristic protocol for combinatorial optimization based on subspace search and quantum circuit sampling.

The protocol consists of two fundamental blocks: selecting an appropriate subspace of the problem and sampling the low-energy states of that subspace. This process (steps ii through v below) is repeated for a predefined number of iterations or until convergence is reached, thereby establishing a descending upper bound for the global optimum at each iteration. Below we detail the steps of the algorithm, illustrated in Fig.~\ref{fig:algorithm_schematic} and condensed as pseudocode in Algorithm~\ref{alg:qssd}. Given a combinatorial optimization problem of $m$ spin variables, let us denote the set of all possible configurations by $\mathcal{S}=\{\mathbf{z}_1,\mathbf{z}_2,...,\mathbf{z}_{2^m}\}$ with $\mathbf{z}_i = (z_i^{(1)},z_i^{(2)},...,z_i^{(m)}) \in \{-1,+1\}^m$, and the lowest-energy configuration found in iteration $t$ of the algorithm by  $\mathbf{z}_t^*$. 

\textbf{i) Initialization.} The initial solution, $\mathbf{z}_0^*$, is set to the all-ones vector, $\mathbf{1}^{m}$. Initially, the set of solutions that are used in the variable selection step is empty, $\mathcal{A}_{0} = \emptyset$. 

\textbf{ii) Variable selection.} At iteration $t > 1$, the set $\mathcal{A}_{t-1}\subset\mathcal{S}$ consists of the $P\in \mathbb{Z}^+$ samples with the lowest energy obtained from iterations $1, 2, ..., t-1$. We refer to $P$ as population size. These $P$ configurations are used to compute a selection score $s_t\in\mathbb{R}$ for each of the problem's variables labeled with an index $v \in \mathbb{Z}^+ \leq m$,
$$
s_t^{(v)} =  \left|\sum_{\mathbf{z}_n\in \mathcal{A}_{t-1}} \frac{z_n^{(v)}}{P}\right| + \frac{y_t^{(v)} - \min(\mathbf{y}_t)}{\max(\mathbf{y}_t) - \min(\mathbf{y}_t)},
$$
that allows each variable to be sorted from highest to lowest relevance. The first term, equivalent to the criterion used in~\cite{atobe_subqubo_2022}, represents the expected value of each spin given the selected samples in $\mathcal{A}_{t-1}$. Variables with a high expected value might have strong local fields or low connectivity, at least considering the information provided by the lower-energy samples obtained up to that point, making them easier to solve locally and thus may not be selected for optimization. The second term tracks the number of times each variable is selected, $\mathbf{y}_t = (y_t^{(1)}, y_t^{(2)}, ..., y_t^{(m)})$, so that the whole solution space could potentially be explored and making the algorithm less likely to get stuck in local minima. Additionally, the second term is scaled to a $[0, 1]$ range, so it has the same range as the first term.  This score is used to sort the variables, $v_i$ with $i=1,\dots,m$, from most to least relevant, that is, from lowest to highest score value, so that $s_t^{(v_i)}\leq s_t^{(v_{i+1})}$.
For the case of $t=1$, when no solutions have been sampled and thus $\mathcal{A}_{t=0}=\emptyset$, the selection score cannot be computed, and so the variables are randomly sorted.

With this calculated score, a subset $\mathcal{V}_{t} = \{v_1, v_2, \dots, v_{n}\}  \subseteq \{1,\dots,m\}$ of exactly $n$ variables is selected. Equivalently, $\mathcal{V}_{t}$ is the lowest-scoring set of $n$ variables that induces a subgraph of the larger problem trying to avoid isolated nodes, considering both quadratic and cubic interactions. Once the set of selected variables is obtained, the values of the rest $\overline{\mathcal{V}}_{t} =
\{1,\dots m\}\setminus\mathcal{V}_{t}=\{v_{n+1}, \dots, v_{m}\}$ are fixed to the spins
$z_{v\in\overline{\mathcal{V}}_{t}} \in \{-1,+1\}$ taken from  $\mathbf{z}_{t-1}^*$. 
The reduced Hamiltonian acting on $\mathcal{V}_{t}$ is
\begin{equation}
\label{eq:H_eff}
\begin{split}
H'_{t}(\mathbf{Z})
=& \ E^{(t)}_0
+ \sum_{i\in\mathcal{V}_{t}} h_{i}^{(t)} Z_{i} 
+ \sum_{\{i,j\}\subset\mathcal{V}_{t}} J_{ij}^{(t)} Z_{i} Z_{j} \\
&+ \sum_{\{i,j,k\}\subset\mathcal{V}_{t}} K^{(t)}_{ijk} Z_{i} Z_{j} Z_{k},
\end{split}
\end{equation}

where

\begin{align}
    h^{(t)}_{i} &= h_{i} + \sum_{a \in \overline{\mathcal{V}}_{t}} J_{ia} z_a + \sum_{\substack{\{a,b\}\subset \overline{\mathcal{V}}_{t} \\ a \ne b}} K_{iab} z_a z_b, \\
    J^{(t)}_{ij} &= J_{ij} +\sum_{a \in \overline{\mathcal{V}}_{t}} K_{ija} z_a, \\
    K^{(t)}_{ijk} &= K_{ijk},
\end{align}
are the reduced coefficients and the constant offset, which is independent of the optimization variables, is obtained as
\begin{equation}
\begin{split}
E^{(t)}_0 =& \
\sum_{a \in \overline{\mathcal{V}}_{t}} h_{a} z_{a}
+\sum_{\substack{\{a,b\}\subset \overline{\mathcal{V}}_{t} \\ a \ne b}}J_{ab} z_{a} z_{b} \\
&+\sum_{\{a,b,c\}\subset \overline{\mathcal{V}}_{t}}
K_{abc} z_{a} z_{b} z_{c} .
\end{split}
\end{equation}

\textbf{iii) Sample subspace.}
Once the reduced subproblem, $H'_{t}$, is defined, the algorithm needs low-energy samples of such subspace. The $S$ samples of iteration $t$, completed with the spins of $\mathbf{z}^*_{t-1}$ on $\overline{\mathcal{V}}_t$, form the set $\mathcal{Q}_t$. By obtaining, not only the optimal solution of the subproblem, but also other quality samples that may lie in a different basin of the energy landscape than the optimal solution, the algorithm might escape local minima and trend towards the global optimum. In principle, any sampling method could work in this framework. Uniform random sampling yields the optimal solution of the subproblem with a probability of $1-(1-2^{-n})^S$, with $S$ being the number of samples. As the value of $n$ increases, the number of samples $S$ needed for the random sampler to obtain good solutions increases exponentially, making the random sampler an inefficient way to explore the energy landscape. Another option is to sample all $2^{n}$ possible solutions, which becomes intractable for  $n\gtrsim 30$. In this work, we propose the use of the QAOA as a particularly suitable tool for this task, since it has been proven to exponentially increase the probability of sampling lower-energy solutions~\cite{diezvalle_qaoa_2023}.

\textbf{iv) Post-process samples.} After obtaining the $S$ samples, a bitflip post-process is applied to the $S_p\in \mathbb{Z}^+ \leq S$ lowest-energy samples obtained. This is implemented by visiting each sample spin once, in a random order, flipping the value, and keeping the change if it results in a lower energy solution than the current configuration. This results in a computational cost that scales as $\mathcal{O}(mS_p)$, which is linear in the number of variables. All of these explored bitstrings, which define the set $\mathcal{B}_t$, are also considered for the variable selection step in the next iteration, being included to select the sample population. This post-processing step is optional, but in practice it has been shown to improve results at a moderate cost.

\textbf{v) Select reference solution.} The set of lowest-energy solutions $\mathcal{A}_{t}$ is updated by keeping the $P$ lowest-energy solutions among $\mathcal{A}_{t-1}$ and the set of all new configurations, considering the ones sampled and those obtained after the post-processing, this is $\mathcal{A}_{t-1} \cup \mathcal{Q}_t \cup \mathcal{B}_{t}$. The best solution so far, $\mathbf{z}_t^* \in \mathcal{A}_{t}$, is set as a reference
for the next iteration $t+1$. 

\textbf{vi) Final solution.} The final output of the algorithm corresponds to the reference solution of the last iteration $T$, $\mathbf{z}_T^*$.

\begin{algorithm}[t]
\caption{Quantum Subspace Sampling Descent (QSSD)}
\label{alg:qssd}
\begin{algorithmic}[1]
\Require cost Hamiltonian $H$ on $m$ spins, number of qubits $n$, iterations $T$, layers $p$, number of samples $S$, population size $P$, post-processed samples $S_p$
\Ensure approximate solution $\mathbf{z}_T^*$
\State $\mathbf{z}_0^* \gets \mathbf{1}^{m}$, \quad $\mathcal{A}_{0}=\emptyset$, \quad $ y_1^{(v)} \gets 0$ $\;\forall v$
\For{$t = 1, \dots, T$}
    \If{$t = 1$}
        \State sort the variables randomly
    \Else
        \State compute $s_{t}^{(v)}$ from the $P$ best strings in $\mathcal{A}_{t-1}$
        \State sort the variables by increasing score
    \EndIf
    \State $\mathcal{V}_{t} \gets$ first $n$ variables with no isolated nodes
    \State $y_{t+1}^{(v)} \gets y_t^{(v)} + 1$ $\forall v \in \mathcal{V}_{t}$
    \State $H'_{t} \gets H\big|_{z_a = z^*_{t-1,a},\ a \in \overline{\mathcal{V}}_{t}}$ 
    \State $(\boldsymbol{\gamma}_t,\boldsymbol{\beta}_t) \gets (\boldsymbol{\gamma}_\infty/A_t,\ \boldsymbol{\beta}_\infty)$, for $p$ layers
    \State $\mathcal{Q}_t \gets \{(\mathbf{x}, \mathbf{z}^*_{t-1}|_{\overline{\mathcal{V}}_{t}})\}$ for $S$ samples $\mathbf{x}$ of QAOA$(\boldsymbol{\gamma}_t,\boldsymbol{\beta}_t)$ on $H'_{t}$
    \State $\mathcal{B}_{t} \gets$ post-process the $S_p$ lowest-energy samples
    \State $\mathcal{A}_{t} \gets P$ best samples in $\mathcal{A}_{t-1} \cup \mathcal{Q}_t \cup \mathcal{B}_{t}$
    \State $\mathbf{z}_t^* \gets \argmin_{\mathbf{z} \in \mathcal{A}_{t}} H(\mathbf{z})$
\EndFor
\State \Return $\mathbf{z}_T^*$
\end{algorithmic}
\end{algorithm}

% =====================

\subsection{QAOA as a non-variational sampler}
\label{sec:qaoa_sampler}

The Quantum Approximate Optimization Algorithm (QAOA) is an algorithm proposed for approximately solving combinatorial optimization problems~\cite{farhi2014quantumapproximateoptimizationalgorithm}. QAOA constructs a parametrized trial state by applying $p$ alternating layers of two non-commuting unitaries to an initial uniform superposition $\lvert + \rangle^{\otimes n}$,
\begin{equation}
  \lvert \psi(\boldsymbol{\gamma}, \boldsymbol{\beta}) \rangle
  = \prod_{l=1}^{p} e^{-i\beta_l H_B}\, e^{-i\gamma_l H_C}
  \, \lvert + \rangle^{\otimes n},
  \label{eq:qaoa-state}
\end{equation}
where $H_C$ and $H_B$ are the cost and mixing Hamiltonians, respectively, and $(\boldsymbol{\gamma}, \boldsymbol{\beta})$ are $2p$ real tunable parameters that control the duration of each evolution. $H_C$ encodes the problem and drives phase separation among candidate solutions based on its cost. At the same time, $H_B = \sum_{j=1}^{n} X_j$, whose ground state is $\lvert + \rangle^{\otimes n}$, mixes amplitude across the computational basis. Candidate solutions are obtained by sampling the state~\eqref{eq:qaoa-state}, with the best-measured bitstring taken as the approximate solution for the combinatorial optimization problem encoded in $H_C$. 

However, in the QSSD framework, QAOA serves as a non-variational low-energy sampler rather than an optimizer. The justification is that even shallow QAOA circuits produce distributions with the structure and properties we need. In particular, for Ising models, the ansatz prepares pure states whose amplitudes in the computational basis follow a Boltzmann-like law~\cite{diezvalle_qaoa_2023, diezvalle_2024},
\begin{equation}
    p(\boldsymbol{z}) = \lvert \langle \boldsymbol{z} \vert \psi(\boldsymbol{\gamma}, \boldsymbol{\beta}) \rangle \rvert^2 \sim e^{-E_{\boldsymbol{z}}/\tau}\;,
    \label{eq:pseudo-boltzmann}
\end{equation}
where $E_{\boldsymbol{z}}$ is the cost of spin configuration $\boldsymbol{z}$, and $\tau$ plays the role of a temperature set by the angles $(\boldsymbol{\gamma}, \boldsymbol{\beta})$ and the number of layers $p$. Depth is a knob that controls the capabilities of the sampler, and in our setting it must be traded directly against circuit fidelity on hardware, as shown in Appendix~\ref{app:layers}. More depth allows for a lower temperature $\tau$ and, therefore, a greater increase in the probability of obtaining not only the global minimum but also low-energy states~\cite{lotshaw_boltzmann_2023,diezvalle_universal_2026}. While proven in the literature for Ising models, we have empirically checked that similar distributions also arise in the HUBO instances tested.    

Two features of this behavior are important for our protocol. First, what the circuit delivers is a population of low-energy configurations rather than a single optimum. This is precisely the resource of QSSD, since the low-energy population both drives the descent and determines the variable-selection scores of the next iteration. Second, the Boltzmann-like character is robust across a wide range of QAOA angles and is not a fine-tuned property of one point in parameter space. Angles that fall short of optimality are thus not a failure mode but a displacement along a family of thermal samplers where the penalty is a warmer distribution, not a qualitative loss of the low-energy bias.

We consequently do not attempt to locate the optimal angles, avoiding the costly variational search of parameters. We instead fix them analytically for each reduced instance following the parameter-setting rules of Ref.~\cite{Sureshbabu2024parametersettingin}, which show that the QAOA energy landscape of a weighted problem approaches that of its unweighted counterpart under a rescaling of $\gamma$ by the root-mean-square magnitude of the coefficients, and that the QAOA objective concentrates sharply around its expectation over random weights. This rule is known to extend to objectives of degree higher than two~\cite{Sureshbabu2024parametersettingin}, which is what allows us to apply it to the HUBO cost functions considered in this work. Concretely, the optimized angles for the Sherrington--Kirkpatrick model in the infinite-size limit, $(\boldsymbol{\gamma}_\infty, \boldsymbol{\beta}_\infty)$ taken from~\cite{basso_quantum_2022}, are rescaled as
\begin{equation}
\label{eq:parameters}
\begin{split}
   \boldsymbol{\gamma}_{t} &= \frac{\boldsymbol{\gamma}_\infty}{A_{t}} \\
    \boldsymbol{\beta}_{t} &= \boldsymbol{\beta}_\infty  
\end{split}
\end{equation}
where
\begin{equation}
\label{eq:A}
\begin{split}
      A^2_{t} =& \ \frac{1}{C_h}\sum_{i\in\mathcal{V}_{t}} (h_i^{(t)})^2 + \frac{1}{C_J}\sum_{{\{i,j\}\subset\mathcal{V}_{t}}}  (J_{ij}^{(t)})^2 \\
      &+ \frac{1}{C_K}\sum_{\{i,j,k\}\subset\mathcal{V}_{t}}  (K_{ijk}^{(t)})^2 
\end{split}
\end{equation}
with $C_h, C_J$ and $C_K$ the number of terms of first, second and third order of $H'_t$, respectively. 
In~\cite{basso_quantum_2022} the $(\boldsymbol{\gamma}_\infty, \boldsymbol{\beta}_\infty)$ vectors for the infinite-size case for $1\leq p\leq17$ are given as optimized values. The maximum number of layers that can be used with this scheme is limited by these available values, although further suboptimal values could be extrapolated from the optimal ones for a higher number of layers.

The rescaling is recomputed at every iteration from the coefficients of the current reduced Hamiltonian, so the sampler adapts to the changing statistics of the subproblems at no quantum and minimal classical cost. The net effect is that a single circuit, of depth growing linearly with $p$, is executed per iteration. In Appendix~\ref{app:landscape}, these angles are shown to reach very similar energy values when compared to the optimal solution for the ideal $p=1$ case. In Appendix~\ref{app:boltzmann}, the distributions obtained for the subproblems are shown to still increase low-energy state sampling probability when using more layers and the analytical angle scheme.

\subsection{Sampler performance}
\label{sec:sampler_performance}

\begin{figure}[t]
\includegraphics[width=0.475\textwidth]{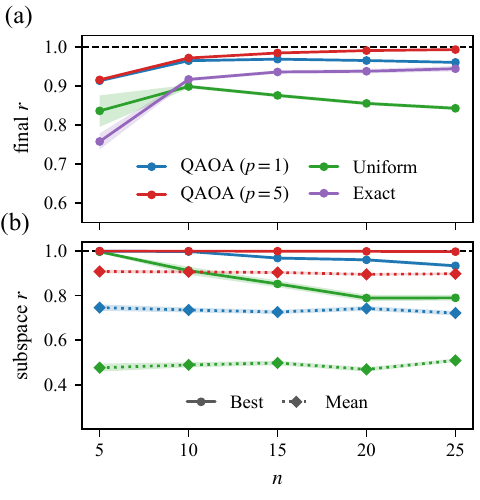}
\caption{Comparison of the performance of QSSD on the heavy-hex ($m=200$) HUBO problem with different samplers: QAOA with 1 (blue) and 5 (red) layers, uniform sampling (green), and the global optimum of each subspace (purple). The dashed black line at $r=1$ marks the global optimum. (a) Comparison of the mean of the algorithm results for the 50 different runs as the number of qubits increases. (b) Per-subspace best (solid line) and average (dotted line) approximation ratio, obtained using the best sample and the mean of all samples respectively, averaged over the subspaces of a single run of 50 iterations. To compute the ratio of each subproblem, Eq.~\eqref{eq:r_scale} is used considering the minimum and maximum energies of the reduced Hamiltonian, $H'_t$, of the subspace. The exact sampler returns a single configuration which always corresponds with $r=1$ and is not shown in (b). Shaded areas represent the 95\% confidence interval of the mean. }
\label{fig:numerical_results}
\end{figure}

In this section, the goal is to evaluate the performance of the QAOA sampler in the context of the full algorithm, compared to a random sampler and to an exact solver that only returns the optimal solution of each subspace. The comparison with the former serves as confirmation that sampling low-energy solutions helps the algorithm converge to a better solution, while the comparison with the latter proves the usefulness of considering more samples over just the optimal solution of the reduced problem. 

For this set of experiments, the QAOA circuit is emulated using an exact state vector emulation and the number of layers is set to $p=1$ and $p=5$, to also assess the impact that the circuit depth has (more details on the depth impact are provided in Appendix~\ref{app:layers}). The QAOA parameters are set as defined in Eq.~\eqref{eq:parameters}. As these emulations can be extremely costly for a large number of qubits, we emulate the behavior of the random and lower-energy sampler for large quantum circuits by sampling a small number of spin strings for smaller values of $n$. A total of $S=100$ strings are sampled in each iteration from the configuration space composed of $2^n$ elements. Furthermore, to isolate the sampler performance of the algorithm, the post-process step is not applied. 

Fig.~\ref{fig:numerical_results} compares the performance of QAOA with $p\in\{1, 5\}$, uniform random sampling and the exact optimal solution on QSSD on the heavy-hex instance with $m=200$ variables, described in more detail in Section~\ref{sec:problem_classes}, for subproblem sizes $n\in\{5,10,15,20,25\}$. The algorithm is run 50 different times, limiting the maximum number of iterations to $T=100$, and setting the population size to $P=500$. Fig.~\ref{fig:numerical_results}a show the final approximation ratios reached by the QSSD algorithm, averaged over the 50 runs, while Fig.~\ref{fig:numerical_results}b represents the mean and best approximation ratios, both averaged over each subspace in a single QSSD run. The reported approximation ratios, $r$, are calculated as
\begin{equation}
\label{eq:r_scale}
    r = \frac{E - E_\mathrm{max}}{E_\mathrm{min} - E_\mathrm{max}}\,,
\end{equation}
where $E_\mathrm{max}$ and $E_\mathrm{min}$ are the highest and lowest energies of the $m=200$ HUBO problem, for the case of Fig.~\ref{fig:numerical_results}a, and in the case of Fig.~\ref{fig:numerical_results}b these are the maximum and minimum energies of each of the reduced Hamiltonian, $H_t'$, of a single QSSD run.

Under these conditions, the QAOA sampler outperforms the exact and random methods at all problem sizes, although some interesting caveats must be borne in mind. Looking at Fig.~\ref{fig:numerical_results}b, the performance gap between QAOA with 5 layers and the random sampler increases with $n$, as the deeper circuit allows QAOA to sample the optimal solution of each subproblem along with other high-quality solutions for all $n$ in almost all cases despite the low number of shots. This is demonstrated by the approximation ratio of the best configuration being at least $r=0.997$, while the mean ratio remains about $r=0.9$. When it comes to the shallower $p=1$ QAOA quantum circuit and the randomly sampled distribution, the gap between the best sample ratio and the optimal of the subproblem at $r=1$ increases with $n$. This plot also shows the importance of setting an appropriate number of layers for the QAOA sampler depending on the number of qubits available. While the average approximation ratio is independent of $n$, the gap between the best sample and the subproblem optimum increases with $n$. This is the reason why, in Fig.~\ref{fig:numerical_results}a, even if for $p=5$ the increased size helps the QAOA explore the complete solution space faster and converge to a better solution with the allowed iteration budget; for $p=1$ and the random sampler, the larger number of qubits is a disadvantage, and the final solution ratio starts dropping for the larger sizes. Using the algorithm with just the optimal solution of each subproblem also yields worse results, even when compared to a shallow QAOA $p=1$ circuit, which often did not sample the optimal solution for the larger sizes tested. The larger pool of samples allows the algorithm to reach a better solution. Even if the difference between the QAOA and exact sampler performance decreases with problem size, using the exact branch-and-bound solver for larger problem instances becomes exponentially expensive and thus infeasible to run.

\subsection{Results on quantum hardware}
\label{sec:hardware_results}

The capabilities of the algorithm on current quantum hardware are benchmarked over the problems defined in Section~\ref{sec:problem_classes} and represented in Fig.~\ref{fig:topologies}, and compared to the classical and quantum solvers listed in Section~\ref{sec:classical_quantum_baseline}. The runtimes of the classical baselines on these instances are included in Appendix~\ref{app:time_cost}. Furthermore, the analysis is extended to explore how future improvements in QPUs in terms of higher qubit counts, lower noise levels and better connectivity could result in better approximate solutions and enable the tackling of more complex problems with a greater number of variables using QSSD.

The QPUs used in these experiments are superconducting qubit processors accessed through the IBM Quantum Platform~\cite{ibm_quantum_platform} using Qiskit~\cite{javadi-abhari_quantum_2024}. The majority of the tests are performed in the 156-qubit, heavy-hex topology \texttt{ibm\_boston}, but the more recent 120-qubit square-lattice \texttt{ibm\_miami} is also used to test the effect of the better connectivity provided by this device. More details on the quantum devices are included in Appendix~\ref{app:hardware}.

All runs are done with $S=10^4$ shots, except for the runs on \texttt{ibm\_miami} which use 1024, since its reported maximum circuits per second (MCPS) is about an order of magnitude lower (250 Hz, against 3.7–3.8 kHz for \texttt{ibm\_boston}). Also, unless stated otherwise, the population size is set to $P=500$ and the post-process step is applied to the $S_p=100$ lowest-energy samples of each iteration. A suitable number of QAOA layers is determined for each problem class based on the tests included in Appendix~\ref{app:layers}.

For problem instances where both $E_\mathrm{min}$ and $E_\mathrm{max}$ are available, the approximation ratio is computed using Eq.~\eqref{eq:r_scale}. Instead, when only a best-known solution with no guarantee of optimality, $E_\mathrm{best}$, is available, the approximation ratio is calculated as 
\begin{equation}
\label{eq:r_min}
    r = \frac{E}{E_\mathrm{best}}.
\end{equation}

\subsubsection{Results on heavy-hex}

The algorithm is first tested on the HUBO problems with heavy-hex topology, $m=200$, and no random cubic terms (see Fig.~\ref{fig:topologies}a). The QSSD algorithm is run 5 times with a maximum number of 50 iterations with $n\in\{ 42, 66, 84\}$, and using both the QAOA and the uniform sampler. The results for these runs are included in Table~\ref{tab:ibm_boston_HUBO_M200} and the evolution of every run is shown in Appendix~\ref{app:hw_runs}. QSSD with $n=42$ and $n=66$ reaches the global optimum in all 5 runs, while the larger and deeper circuit of $n=84$ struggles a bit more, failing to reach the optimum in 3 out of the 5 runs, but making it there faster when it does not get stuck. One possible explanation for this is the fact that reaching the optimal solution of each subspace after the post-process is very likely for $n=42$ and $n=66$, but is harder for $n=84$. This is the same property seen in Section~\ref{sec:sampler_performance}, as $n$ increases, more layers $p$ are needed in order for the subproblem optimum to be sampled. Due to the structure of the couplings between the variables, these instances are particularly well suited for exact solvers such as HiGHS, which are capable of finding the global optimum in a short amount of time. The SA algorithm implemented, detailed in Section~\ref{sec:classical_quantum_baseline}, also reaches the global optimum with $10^4$ reads and $10^4$ sweeps, and the greedy baseline with $10^5$ random starts stops at $r=0.978$, below every QSSD run.

\begin{figure*}[t]
\centering
\includegraphics[width=1\textwidth]{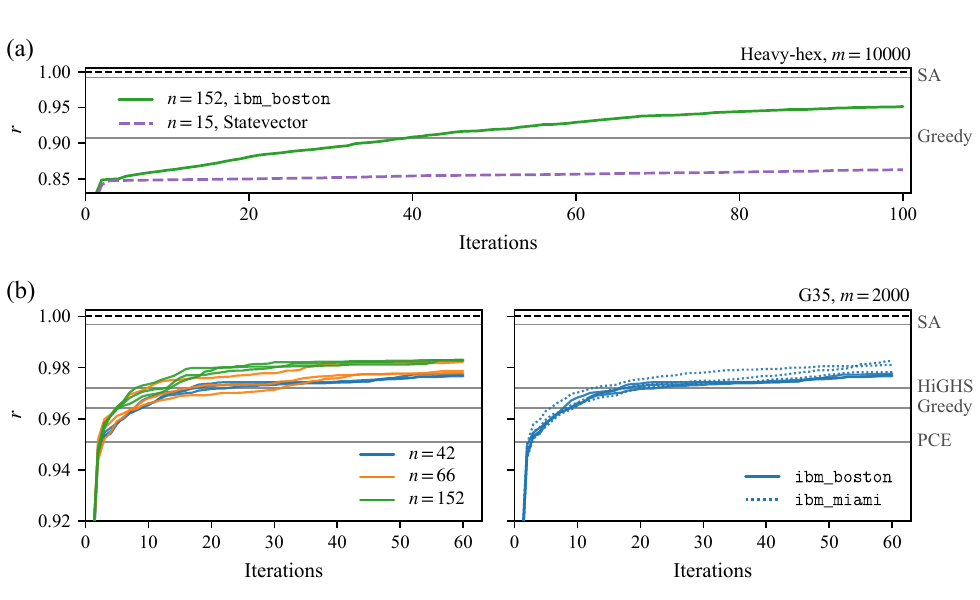}
\caption{Evolution of the approximation ratios of the reference solutions for different QSSD configurations on the heavy-hex with $m=10000$ variables and G35 problem instances. (a) Heavy-hex with $m=10000$, one run is performed with $n=152$ on \texttt{ibm\_boston} and one with $n=15$ using ideal state vector emulation, both with $p=8$. (b) G35 with $p=4$. Left: three runs for each size $n\in\{42,66,152\}$ on \texttt{ibm\_boston}. Right: three runs with $n=42$ on \texttt{ibm\_boston} (solid) and on \texttt{ibm\_miami} (dotted). The dashed black line marks the optimum obtained with HiGHS in (a) and the best-known cut, 7687, in (b), and the gray lines the results of SA, HiGHS, PCE, and the greedy baseline (Table~\ref{tab:gset_g35}).}
\label{fig:iterations_M10000_g35}
\end{figure*}

In order to explore the limits of our approach, the algorithm is also tested on a significantly larger instance with $m=10000$ variables. The optimal solution for the problem, despite its large size, is found in less than an hour with the HiGHS solver, due to its inherent structure, which favors the branch-and-bound algorithm as explained above. The SA implementation reaches $r=0.992$ with $10^4$ reads and $10^4$ sweeps, without finding the optimum, and the greedy baseline reaches $r=0.907$. As shown in Fig.~\ref{fig:iterations_M10000_g35}a, the QSSD algorithm reached $r=0.95$ after 100 iterations, using $n=152$ qubits with 8 QAOA layers on the \texttt{ibm\_boston} backend, and it continues to improve gradually with more iterations. Additionally, even though the ideal state vector with $n=15$ run does not have to deal with noise, it falls to a solution about 10\% worse than the \texttt{ibm\_boston} run, showcasing the fact that using a quantum computer is needed in order to be able to sample from subproblems of larger size $n$ and improve QSSD performance.

\begin{table}[t]
\centering
\small
\renewcommand{\arraystretch}{1.2}
\setlength{\tabcolsep}{0pt}
\begin{tabular*}{\columnwidth}{@{\extracolsep{\fill}} cccccc}
\toprule
\textbf{Sampler} & {$n$} & $p$ & $\bar{r}$ & \textbf{Success} (\%) & \textbf{Iters.} \\
\midrule
QAOA   & $42$ & $5$ & 1.0000 & 100  & $27.2 \pm  10$ \\
QAOA   & $66$ & $5$ & 1.0000 & 100  & $36.2 \pm   4$ \\
QAOA   & $66$ & $7$ & 1.0000 & 100  & $23.6 \pm  14$ \\
QAOA   & $84$ & $9$ & 0.9997 & 40   & $21.5 \pm 1.5$ \\
Uniform & $42$ & -   & 0.9995 & 60   & $25 \pm 15$   \\
Uniform & $66$ & -   & 0.9963 & 0    & -    \\
Uniform & $84$ & -   & 0.9920 & 0    & -    \\
\bottomrule
\end{tabular*}
\caption{Results of the QSSD algorithm using the QAOA, with various circuit depths $p$ on \texttt{ibm\_boston}, and the uniform random samplers, across qubit counts $n$ for the heavy-hex with $m=200$ variables. The metrics included are the mean approximation ratio $\bar{r}$, the percentage of runs that reach the global optimum and the average number of iterations considering only the successful runs.}
\label{tab:ibm_boston_HUBO_M200}
\end{table}

\subsubsection{Results on Gset}

The application of QSSD on instances of the well-known benchmark set Gset yields promising results. The algorithm is tested on the G1 and G35 instances, with $m=800$ and $m=2000$ variables respectively. Due to the nature of the graphs, the HiGHS algorithm struggles to reach the optimal solution with a limited time of 2 hours. Compared to the heavy-hex problems, these Gset instances are much harder for HiGHS. The SA heuristic, however, reaches the best-known solution for G1 and a ratio of $r=0.997$ for G35, with the reported $10^3$ sweeps and $10^3$ reads. The greedy baseline with $10^5$ random starts reaches $r=0.989$ on G1 and $r=0.964$ on G35. Additionally, the reported PCE ratios in~\cite{sciorilli_towards_2025} for these two instances are included in order to have another quantum algorithm for comparison. 

For the case of the G35 problem, QSSD is executed 3 times with 60 iterations for different numbers of qubits and QPU connectivities, using the heavy-hex \texttt{ibm\_boston} with $n\in \{42, 66, 152\}$ and the square-lattice \texttt{ibm\_miami} with $n=42$. All circuits are run with $p=4$ layers, for all sizes and for the two QPUs. These runs are included in Fig.~\ref{fig:iterations_M10000_g35}b and in Table~\ref{tab:gset_g35}. QSSD reaches better solutions than the PCE, the HiGHS algorithm, and than the greedy baseline, although the SA is, in this case, the best option.

It is worth mentioning the increase in both the best and mean ratio as the number of qubits increases in the \texttt{ibm\_boston} QPU. Both the highest mean and overall result are obtained using $n=152$ qubits. Additionally, QSSD is run for smaller sizes, $n\in\{15,  20\}$, with ideal state vector emulation of the QAOA sampler, setting the number of shots to $10^4$. Even when comparing the results for the emulated runs using a similar number of qubits as PCE, QSSD reaches better solutions after a few iterations. It should also be noted that runs on \texttt{ibm\_miami} achieve better approximation ratios than those performed on \texttt{ibm\_boston}. The better connectivity of the device allows for improved quantum circuit transpilation and reduced noise, which demonstrates how improvements in quantum hardware will lead to better algorithm performance.

\begin{table}[t]
\centering
\small
\renewcommand{\arraystretch}{1.2}
\setlength{\tabcolsep}{2.5pt}
\begin{tabular}{llcc}
\toprule
\textbf{Solver} & \textbf{Setting} & \textbf{Best $r$} & \textbf{Mean $r$} \\
\midrule
\multirow{6}{*}{QSSD }
 & $n=15$, statevector &  0.9792  & $0.9763 \pm 0.0020$ \\
 & $n=20$, statevector &  0.9772  & $0.9754 \pm 0.0018$ \\
 & $n=42$, \texttt{ibm\_boston} &  0.9776  & $0.9771 \pm 0.0013$ \\
 & $n=42$, \texttt{ibm\_miami} &  0.9826  & $0.9806 \pm 0.0018$ \\
 & $n=66$, \texttt{ibm\_boston} &  0.9823  & $0.9796 \pm 0.0060$ \\
 & $n=152$, \texttt{ibm\_boston} &  0.9831  & $0.9829 \pm 0.0005$ \\
\midrule
PCE~\cite{sciorilli_towards_2025} & $n=17$ & 0.951 & - \\
SA & $10^3$ sweeps/reads & $0.9969$ & - \\
HiGHS & 2\,h max. time & $0.9720$ & - \\
Greedy & $10^5$ starts & $0.9641$ & - \\
\bottomrule
\end{tabular}
\caption{Best and mean approximation ratio, $r$, results on Gset instance G35 ($m=2000$). QSSD reports the mean $\pm$ 95\% confidence interval of the best cut found within 60 iterations, for circuit depth $p=4$, across qubit counts $n$. PCE, simulated annealing (SA), greedy local search and HiGHS are reported as single-value baselines each from a single run. The approximation ratio is the cut value normalized by the best-known cut ($7687$)~\cite{ma2017multiple}.}
\label{tab:gset_g35}
\end{table}

For the smaller G1 instance, the QSSD algorithm is run once on \texttt{ibm\_boston} using $n=42$ and $p=2$ for 50 iterations to reach a ratio of $r=0.987$, which is higher than the HiGHS solver with the 2-hour time limit and the reported PCE ratio. The full run is included in Appendix~\ref{app:hw_runs}. Again, the SA is a good heuristic for this problem, reaching the best-known solution with the allowed number of reads and sweeps. Furthermore, the greedy baseline, with $r=0.989$, is also slightly above the QSSD hardware run in this case.

\subsubsection{Results on heavy-hex with random cubic terms}

The results of the two previous problem types, heavy-hex HUBO problems and the QUBO Gset instances, show that, while QSSD can perform reasonably well for both, there is a better classical alternative. For the heavy-hex HUBOs, the HiGHS solver can take advantage of the sparsity and structure, finding the guaranteed optimum very efficiently for the tested problem sizes, while for the G1 and G35 problem instances, a standard SA and even the greedy baseline in the smaller G1 instance reach better solutions. Motivated by these findings, a new problem type is defined to be hard for HiGHS, while being sparse enough so that the resulting QAOA circuits do not explode in depth and gate count. These new problems consist of taking the same heavy-hex HUBO problems and include cubic terms that are not restricted to the topology with a very low probability of $\rho_C = 5\cdot10^{-5}$ (see Fig.~\ref{fig:topologies}c). This means that for every possible unique combination of three variables (considering $(i, j, k)$ equal to $(j, i, k)$ and all other permutations) a coefficient $K_{ijk}$ is sampled from the standard normal distribution $\mathcal{N}(0,1)$ with probability $\rho_C$. This results in the following expected number of non-structured cubic coefficients,
\begin{equation}
    \mathbb{E}[C_K^r] = \frac{m(m-1)(m-2)}{3!}\rho_C,
\end{equation}
where $C_K^r$ denotes the number of random cubic terms added on top of the heavy-hex native ones. 

For these problem instances, the HiGHS solver, which is the solver used to obtain the optimal solution to serve as reference in previous sections, could not converge to a 0\% optimality gap for the larger problems, $m\in\{400, 500\}$ given the set limit of 72 hours. Thus, the approximation ratios in Table~\ref{tab:hubo_summary} are obtained following Eq.~\eqref{eq:r_min} and the best-known energy is set to the lowest energy obtained by any of the algorithms, which for $m\in\{100, 200, 300\}$ is the optimal solution found with HiGHS, while for $m\in\{400, 500\}$ is the best SA solution. The value included in the table for the HiGHS algorithm for $m\in\{400, 500\}$ is the value for the best solution achieved up to the point of the timeout. The evolution of $r$ as a function of the number of iterations for each size is reported in Appendix~\ref{app:hw_runs}.

\begin{table}[t]
\centering
\small
\renewcommand{\arraystretch}{1.2}
\setlength{\tabcolsep}{3pt}
\begin{tabular}{ll cccc}
\toprule
    & & \multicolumn{4}{c}{\textbf{$m$}} \\
\cmidrule(lr){3-6}
\textbf{Solver} & \textbf{Setting} & 200 & 300 & 400 & 500 \\
\midrule
HiGHS & 72\,h max. time & $1.000$ & $1.000$ & $0.936$ & $0.849$ \\
HiGHS & $10^3$ max. nodes & $1.000$ & $0.839$ & $0.648$ & $0.560$ \\
SA & $10^4$ sweeps/reads & $1.000$ & $1.000$ & $1.000$ & $1.000$ \\
Greedy & $10^5$ starts & $0.934$ & $0.897$ & $0.875$ & $0.865$ \\
QSSD & $n=20$, statevector & $0.968$ & $0.912$ & $0.871$ & $0.888$ \\
QSSD & $n=66$, \texttt{ibm\_boston} & $0.999$ & $0.950$ & $0.934$ & $0.886$ \\
\bottomrule
\end{tabular}
\caption{Solution quality (approximation ratio, $r$, relative to the best-known energy) for each solver on the random heavy-hex HUBO instances of size $m\in\{200,300,400,500\}$ with cubic probability $\rho_C=5\cdot10^{-5}$. The reference energy is the best solution found by all the optimizers, which is the exact HiGHS optimum for $m=200$ and $m=300$, and the SA solution for $m=400$ and $m=500$. HiGHS timed out for $m\in\{400, 500\}$ ($\approx3$ days), so the best solution up to that point is recorded.}
\label{tab:hubo_summary}
\end{table}

QSSD is run once for 50 iterations and using two different sizes, $n=20$ and $n=66$, the former by emulation and the latter using the \texttt{ibm\_boston} backend, with $p=5$ and $p=3$ respectively. QSSD with $n=66$ is above the greedy baseline at every size, and the run with $n=20$ is as well except at $m=400$. Note that, only for $m=500$, the emulated execution $n=20$ finds a marginally better solution than the real hardware one, likely because only a single run is available. SA reaches the best-known energy at every size, and is the best method on these instances.
HiGHS reaches a slightly better non-optimal solution than QSSD for $m=400$, while for $m=500$ both QSSD runs improve on the HiGHS solution by about 4\%.

\subsection{Runtime scaling}
\label{sec:runtime}

\begin{figure*}[t]
\centering
\includegraphics[width=\textwidth]{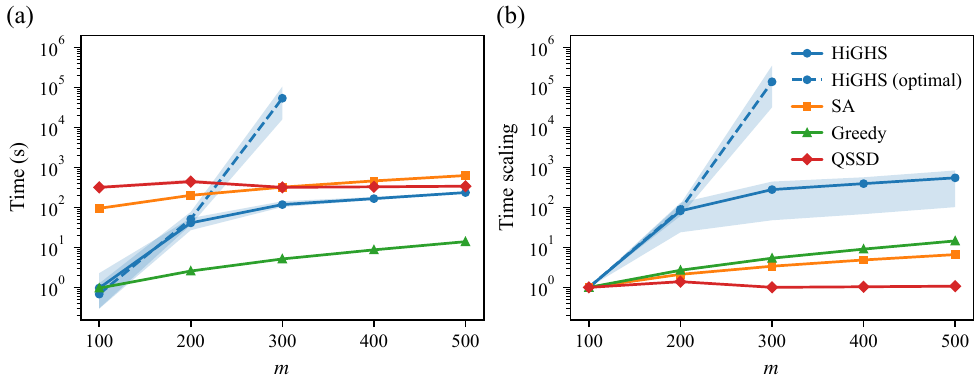}
\caption{(a) Runtime in seconds of HiGHS ($10^3$ nodes), SA ($10^4$ reads $\times$ $10^4$ sweeps), greedy ($10^5$ starts) and QSSD ($T=50$, $n=66$, $p=3$) on heavy-hex HUBOs with random cubic terms ($\rho_C=5\cdot10^{-5}$) as a function of $m$. QSSD time is the measured time of the complete run of $T$ iterations including QPU time per circuit but not the queue time. The dashed line is the time HiGHS needed to prove optimality on the same instances. It is shown only up to $m=300$, the largest size solved for every instance, due to the extremely long runtimes. (b) Time scaling, computed as the time relative to the same method at $m=100$. Lines show the mean over three problem instances per size and shaded areas the minimum-maximum range.}
\label{fig:time_scale_fixed}
\end{figure*}

Beyond the accuracy, when looking at how the computational cost of each algorithm scales as the problem size increases, the QSSD approach is advantageous. As shown in Fig.~\ref{fig:time_scale_fixed}, the exponential scaling of HiGHS to achieve the global optimum makes it an unsuitable option for large instances. When a maximum number of iterations is set, both the SA and HiGHS scale worse than QSSD for a fixed $n$. 

The runtimes are compared for a fixed amount of resources per method, independent of $m$. Three heavy-hex HUBO instances with random cubic terms, generated as explained in Section~\ref{sec:problem_classes} with $\rho_C=5\cdot10^{-5}$, are used for each size $m\in\{100,200,300,400,500\}$, one of them being the ones reported in Table~\ref{tab:hubo_summary}. HiGHS is included in two different configurations, one that is left as much time as possible until the optimal solution is found and then one that is stopped after $10^3$ branch-and-bound nodes. SA runs $10^4$ reads of $10^4$ sweeps, the greedy baseline performs local search on $10^5$ random starts, and QSSD is limited to $T=50$ iterations. The rest of the hyperparameters of QSSD are the same as the ones used to obtain the results in Table~\ref{tab:hubo_summary}: $n=66$, $p=3$, $S=10^4$, $P=500$ and $S_p=100$. All classical computations run on a single pinned core of the same laptop with an Intel Core i7-13650HX CPU. To estimate the QSSD cost for the other instances not run on the quantum hardware, the measured samples are replaced by uniform samples of the same size. This is justified for this test, since only the times are measured and not the quality of the results. The classical steps of an iteration are then timed with these samples, and the quantum part is the QPU time reported by IBM for the $n=66$ circuits of Table~\ref{tab:hubo_summary}, about 5\,s per circuit independently of $m$, which excludes queueing (the time the job waits on the IBM Quantum Platform before it runs, which is very variable depending on server load).

The results for the total runtime of each algorithm with the given budget, along with the time HiGHS needs to reach optimality (dashed blue line) are shown in Fig.~\ref{fig:time_scale_fixed}a, and the time scaling with problem size with respect to the $m=100$ instance is included in Fig.~\ref{fig:time_scale_fixed}b. The time HiGHS takes to reach the optimal solution grows exponentially with $m$ and was not run for larger $m$, due to the long execution times. With the budget of $10^3$ nodes, most of the smallest instances are solved before the node limit is reached, so the time stays low, and it only starts to grow steadily from $m=300$, when every instance reaches the limit. For SA and the greedy baseline, the work per step is proportional to the number of terms of the problem, so their times grow with $m$, faster for the greedy baseline than for SA, although the greedy baseline remains more than an order of magnitude faster in total runtime. In contrast, QSSD is nearly constant over the range tested, as it is dominated by the QPU time, which depends on $n$, $p$ and $S$ but not on $m$, while its classical part grows about linearly with problem size. These include the variable selection, the reduced Hamiltonian building and the post-process step. Moreover, the transpilation optimization level selected is the highest available (level 3 in Qiskit's preset transpiler), the same one used for all the real hardware circuits in this work. Due to this, the transpilation of a few iterations takes a lot longer than the rest, as shown in the case $m=200$, increasing the mean runtime disproportionately for that size.

\section{Discussion}
\label{sec:discussion}

In this work, QSSD is introduced as an iterative heuristic algorithm for solving combinatorial optimization problems. It uses QAOA with non-variational precomputed angles to sample low-energy configurations of subspaces of the original problem, thus allowing QSSD to tackle instances with more variables than the available number of qubits. As the angles are not optimized variationally, QSSD only requires one quantum circuit execution per iteration.

The algorithm is applied to three different problem classes: heavy-hex HUBO problems, two Max-cut instances from the Gset suite (G1 and G35) and heavy-hex problems with additional random cubic terms. To assess the ability of QSSD to reach good solutions, three classical algorithms: the branch-and-bound solver HiGHS, simulated annealing (SA), and multistart greedy local search, are also applied to the tested problem instances. For the G1 and G35 problems, the reported results of the Pauli Correlation Encoding (PCE) algorithm, a quantum algorithm that also reduces the necessary number of qubits, are included as well. 

For almost all problem classes and instances, QSSD is proven to be able to obtain good solutions, outperforming the greedy baseline and even HiGHS for the more complex instances, but it still falls short of the simulated annealing algorithm within the number of QSSD iterations tested. Even in comparison to the latter, the results for the heavy-hex with random cubic terms instances show that SA (and the other classical baselines tested in this work) scale worse in runtime than QSSD for executions with a fixed iteration budget. Therefore, as problem size keeps increasing, there is a configuration regime (number of qubits, layers and iterations) where potentially QSSD could be preferred, though that remains to be proven in future work.

Throughout these tests, the use of QAOA as a sampler in QSSD, even with non-optimal angles, is shown to outperform uniform random sampling for a large enough $n$ and $p$, as the solution quality reached with uniform samples drops for large $n$ (Appendix~\ref{app:samplers}). Considering just the optimal solution of the subproblems also performs worse than QAOA for the relevant range for QSSD, when $n \ll m$, as it benefits from a larger number of explored configurations. For larger $n$ and $m$, the cost of finding the optimal solution of each subproblem grows exponentially, rendering the exact solver inefficient in that context.

\begin{figure}[t]
\centering
\includegraphics[width=\columnwidth]{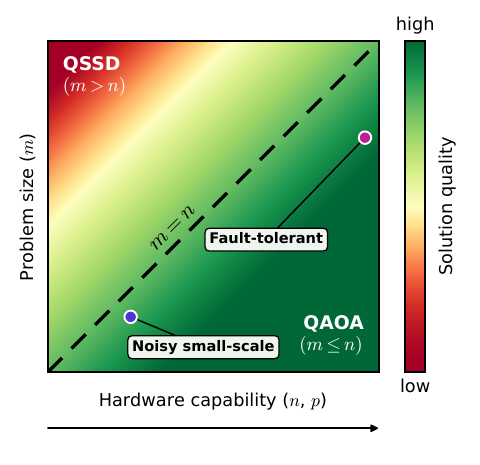}
\caption{Regimes of QSSD solution quality as a function of the hardware capability (considering the available number of qubits $n$ and circuit depth that can be run without noise interference, $p$), and of the problem size, $m$. QAOA can be applied to problems that fit on the device, $m\le n$, and the quality of the solution will depend on circuit depth. QSSD targets problems with more variables than qubits, $m>n$, where a direct QAOA approach is not feasible, but can still take advantage of improvements in hardware capability that increase QAOA solution quality.}
\label{fig:qssd_regime}
\end{figure}

The main feature of QSSD is the decoupling of the number of variables, $m$, and the number of qubits, $n$, which allows it to be applied to problems with more variables than the currently available number of qubits, while still benefiting from future quantum hardware improvements. Figure~\ref{fig:qssd_regime} represents this idea, by showing the regime where QSSD is useful ($m > n$), but also how, as the depth and size of the quantum circuits that can be run on the quantum processor increase, so does the performance of the QAOA sampler and thus the performance of QSSD. Nevertheless, just increasing $n$ is not enough, as $p$ must increase too to maintain the sampler performance, as shown in the results of Fig.~\ref{fig:numerical_results}a, where the one-layer QAOA performance increases with $n$ up to a point where it starts to degrade. To be able to run deeper circuits without being limited by noise, better hardware fidelity, but also better connectivity and compilation strategies are needed. This is consistent with the improvement in performance going from the heavy-hex topology of the \texttt{ibm\_boston} to the square-lattice topology of the \texttt{ibm\_miami} for the G35 instance, where the shallower transpiled circuits can provide higher-quality samples for the same number of layers. 

In the limit $n=m$, QSSD just becomes plain QAOA with precomputed angles and a post-process. Therefore, in the limit of infinite number of layers, $p\rightarrow \infty$, with no errors and an adiabatic-like angle schedule it is guaranteed to reach the optimal solution in a single iteration with a feasible number of shots~\cite{farhi2014quantumapproximateoptimizationalgorithm}. This also highlights the fact that the ability of QSSD to reach the global optimum of the optimization problem is determined by the ability of the sampler plus the post-process to find it once it belongs to the sampled subspace. Furthermore, the lower the probability of sampling the optimal solution (and close neighbors) of a given subproblem, the higher the number of iterations the algorithm will need to converge to the global optimum of the complete problem. However, once the optimum is sampled at a sufficiently high rate, increasing this probability up to 100\% again hurts QSSD performance, as seen in the sampler performance results using the exact solver in Section~\ref{sec:sampler_performance}.  

The reliance on the sampler also presents some limitations on the current implementation and opens avenues to improve the algorithm. First, the performance of the QAOA sampler is bounded by the fixed-angle scheme used, which is shown to be good enough to increase low-energy sample probability. Nevertheless, there is room for improvement on the current scheme. Alternatively, other sampling algorithms, such as a greedy search or SA, could be tested in place of QAOA. Second, the subspaces selected by QSSD are in general fragmented and not native to the device. Their variables form several disconnected pieces of the problem graph and acquire couplings outside the lattice of the quantum computer topology through the frozen terms, which requires SWAP gates. As a result, the dense G1 instance could only be run with $n=42$ and few layers, $p=2$, in the heavy-hex \texttt{ibm\_boston} device. Both a better variable selection strategy and a more general transpilation procedure could reduce circuit depth and its sensitivity to noise, allowing for larger $n$ and $p$ to be used. Third, the hardware results rely on few runs per setting, since executions on quantum hardware are expensive. Hence, the characterization of the algorithm performance and its properties could benefit from additional runs. 

Aside from the algorithm itself, there are challenges to further assess the capabilities of QSSD to solve large-scale optimization problems. The reported QSSD runtimes include the QPU time per circuit but not the queueing times of the cloud access to the devices, which can dominate the wall-clock time of a run and could be a real drawback for a production-ready implementation of the algorithm. Moreover, the comparison with classical methods is limited to an exact solver, a trivial greedy local search and simulated annealing, hence better classical heuristics may exist for some of the instances considered. Note that the goal of the work is not to prove quantum advantage and claim that QSSD outperforms the best classical algorithms, but rather to show that QSSD can fill a niche to reach approximate solutions for instances and sizes where exact solvers such as HiGHS scale poorly, while at the same time being able to harness the potential of QAOA and improving with the size and fidelity of the quantum hardware. 

Even considering the caveats and potential improvements discussed above, the structure of QSSD allows current quantum hardware to address instances with thousands of variables, and every gain in qubit count, fidelity or connectivity directly translates into larger subproblems, deeper circuits and better solutions, without changes to the protocol. QSSD is thus a practical heuristic for combinatorial optimization problems that helps bring quantum optimization a step closer to utility.

\section{Methods}
\label{sec:methods}

\subsection{Problem classes}
\label{sec:problem_classes}

The problems considered in this work belong to two main classes, higher-order unconstrained binary optimization (HUBO) problems and Max-cut. A HUBO problem defined over binary variables can be easily transformed to a cost function of $m$ spin variables $\mathbf{z}\in\{-1,+1\}^m$ of the form in Eq.~\eqref{eq:hubo}. For cost functions of degree three, as the ones used here, the problem Hamiltonian takes the form of Eq.~\eqref{eq:hubo}, with linear, quadratic and cubic coefficients $h_i$, $J_{ij}$ and $K_{ijk}$. When all cubic coefficients vanish, the problem reduces to a Quadratic Unconstrained Binary Optimization (QUBO) problem, equivalent to an Ising model. A HUBO can also be rewritten as a QUBO, but only by adding auxiliary variables and penalty terms, which increases both the number of variables and the density of the interactions. In QSSD, as in QAOA, the higher-order terms are instead implemented directly in the circuit.

The Max-cut problem is defined on a graph with edge weights $w_{ij}$, with $w_{ij}=0$ if vertices $i$ and $j$ are not connected. The task is to divide the vertices into two sets such that the total weight of the edges connecting the two sets, the cut, is maximal. Assigning to each vertex a spin $z_i=\pm1$ that indicates its set, the cut value is
\begin{equation}
    C(\mathbf{z}) = \sum_{i>j} w_{ij}\,\frac{1-z_iz_j}{2}\,,
\end{equation}
and maximizing $C$ is equivalent to minimizing $E(\mathbf{z})=-C(\mathbf{z})$. This is a QUBO problem of the form of Eq.~\eqref{eq:hubo}, with $h_i=0$, $J_{ij}=w_{ij}/2$ and $K_{ijk}=0$, up to a constant offset.

\begin{figure*}[t]
\centering
\includegraphics[width=\textwidth]{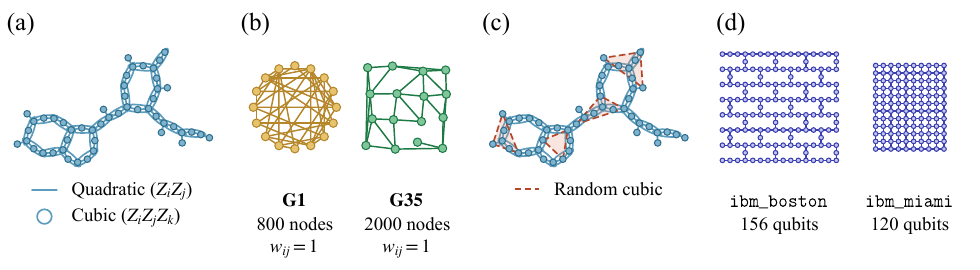}
\caption{Representation of the topologies of the three problem classes tested and of the quantum devices used. (a) Representation of a heavy-hex HUBO, edges are quadratic terms and the shaded ellipses are the cubic terms supported on three neighboring variables. (b) Representations of the Gset Max-cut instances G1 and G35. The G1 instance is a random graph, while the G35 is planar, and both have all weight values, $w_{ij}=1\;\forall i,j$. (c) The same problem topologies as (a) with added random cubic terms, which can involve any three variables, not just neighboring nodes, represented as dashed triangles. (d) Coupling maps of \texttt{ibm\_boston} (heavy-hex, 156 qubits) and \texttt{ibm\_miami} (square lattice, 120 qubits).}
\label{fig:topologies}
\end{figure*}

From these two families, three problem classes are built: heavy-hex HUBO problems, Max-cut instances of the Gset suite, and heavy-hex HUBO problems with additional random cubic terms. A representation of the topologies of the problem classes and the quantum devices is shown in Fig.~\ref{fig:topologies}.

\textbf{i) Heavy-hex.} The heavy-hex problems are built based on a random heavy-hex topology, the same class of topology as the IBM Heron quantum computers, which are the main devices used throughout this work, although not exclusively. This ensures the resulting QAOA circuit has low depth when run on this kind of backends by being able to compile the quadratic and cubic terms as shown in~\cite{pelofske_short-depth_2024}. As the topology is not the exact same as the hardware, the transpilation process has to insert SWAP gates. The coefficients of the HUBO problems, defined in Eq.~\eqref{eq:hubo}, are defined as $J_{ij} = w_{ij}/4$ and $h_i = - w_{ii}/2 - \sum_{j \in N_i} w_{ij} / 4 $, where $N_i$ denotes the neighbors of variable $i$ in the heavy-hex lattice, while $w_{ij}$ and $K_{ijk}$ are drawn from a normal distribution with zero mean and unit variance, $\mathcal{N}(0,1)$. Building the HUBO Hamiltonians with this topology in mind allows us to optimize the QAOA circuit depths, by implementing the quadratic and cubic terms in the circuit as shown in~\cite{weidenfeller_scaling_2022}.

\textbf{ii) Gset.} The Gset is a common benchmark suite~\cite{helmberg_spectral_2000, ye_gset_nodate} designed to test algorithms on combinatorial optimization problems. The set includes 71 Max-cut graphs of different sizes and densities. In this paper, the algorithm is tested on the G1 and G35 instances, which have 800 and 2000 variables respectively, and the results are compared to the best-known solutions~\cite{ma2017multiple}, and the solutions obtained using other classical optimizers. G1 is a random graph with $m=800$ vertices and 19176 edges (density of 6\% and 48 average degree), while G35 is a less dense graph with $m=2000$ vertices and 11778 edges (density of 0.6\% and 11.8 average degree). In both instances all the edge weights are $w_{ij}=1$, making them unweighted Max-cut problems.

\textbf{iii) Heavy-hex with random cubic terms.} To increase the complexity of the combinatorial problem, the algorithm is tested on instances of sizes $m\in\{100, 200, 300, 400, 500\}$ with extra cubic terms that do not follow the topology. An extra cubic term, $K_{ijk}$ with $i>j>k$, is included with a probability $\rho_C=5\cdot10^{-5}$ and is again sampled from $\mathcal{N}(0,1)$. These extra terms break the structure of the heavy-hex problem, making it much harder for the branch-and-bound-based classical solver to reach the optimal solution.

\subsection{Quantum circuit optimization}
\label{sec:circuit_optimization}

The QAOA quantum circuit for each iteration needs to be adapted to the specific quantum hardware. In this work, the compilation strategy introduced in~\cite{weidenfeller_scaling_2022}, is used as a pre-processing step to order the two-qubit gates of the reduced Hamiltonian, $H'_t$ of each iteration. The edge coloring of the interaction graph of $H'_t$ is computed, so that each edge sharing a common node is assigned to a different color, and each color corresponds to a layer of parallel gates. The optimal edge coloring is used when the graph is bipartite, as is the case for the smaller subproblems of the heavy-hex problems, while greedy coloring is used otherwise, such as the Gset instances and for the larger heavy-hex subproblems. With the resulting coloring scheme, the QAOA circuit is built considering neighboring quadratic and cubic interactions as described in~\cite{pelofske_short-depth_2024}. Each cost layer, $e^{-i\gamma_l H'_t}$, is built by first implementing the linear terms as single-qubit rotations, $R_z(2\gamma_l h^{(t)}_i)$, then the quadratic and cubic interactions between neighboring nodes are included using the coloring scheme, and finally the cubic terms that cannot be nested in the coloring scheme are handled separately and appended at the end of the layer. When the cubic term, $K^{(t)}_{ijk}$, involves variables $i,j,k$ with corresponding quadratic terms $J^{(t)}_{ij}$ and $J^{(t)}_{jk}$, it can be nested between the CNOTs of the neighboring quadratic terms without additional CNOT gates. This means that the cost of including these cubic terms is one single-qubit gate, $R_z(2\gamma_l K^{(t)}_{ijk})$. The circuit alternates these cost layers in forward and reversed gate order from one QAOA layer to the next, each followed by the mixer gates, $R_x(-2\beta_l)$. To adapt the circuit to the specific quantum hardware, the Qiskit preset pass manager is used at optimization level 3~\cite{javadi-abhari_quantum_2024}. This step selects the physical layout, inserts the SWAP gates required when the interaction graph is not a subgraph of the device coupling map, and translates the circuit operations to the native gate set of the quantum computer. Additionally, no error mitigation or suppression is applied.

For the ideal case of a hardware-native problem, the interaction graph is a subgraph of the heavy-hex lattices, so the coloring requires at most three colors and the cubic terms can be nested between the CNOTs of the neighboring quadratic terms without additional CNOT gates.

Under these conditions, the cost layer has a maximum CNOT depth of six, independently of the number of qubits and of cubic terms. Nevertheless, the reduced Hamiltonians that QSSD generates do not keep this structure, as the selected variables often come from different sections of the complete graph, and the cubic terms with a frozen variable sometimes may introduce interactions between variables that are not neighbors in the lattice, as is the case when the central variable of a cubic term is frozen while the other two are selected. Despite this, the construction reduces the two-qubit depth of the transpiled circuits of the heavy-hex problems with respect to the standard QAOA circuit, mainly due to the cubic terms sharing the CNOTs of the quadratic terms and the two-qubit color ordering. For the problem types that do not follow the device topology, this improvement is less clear (Appendix~\ref{app:transpilation}).

\subsection{Classical and quantum baselines}
\label{sec:classical_quantum_baseline}

To compare the performance of the QSSD algorithm, the tested problems are also tackled using classical optimization algorithms with different properties, exact and heuristic. In addition, the results reported in the literature of other quantum algorithms are included as a benchmark. 
\begin{enumerate}
    \item The exact solver used to obtain the optimal solution of the problems when possible is the branch-and-bound-based \textbf{HiGHS} solver~\cite{HiGHs}. The implementation solves a linear integer programming problem, so all quadratic and higher-order terms must be reduced to linear terms, at the cost of adding variables and constraints. This solver is fast for problems that have a certain underlying structure, but its runtime increases exponentially with the number of problem variables. 
    \item The simulated annealing (\textbf{SA}) algorithm is a stochastic optimization metaheuristic for global optimization that accepts worse solutions with a certain probability that decreases over the iterations of the algorithm. This strategy allows it to initially explore the search space and escape local minima before reaching a near-optimal solution~\cite{simulated_annealing}. We use a single-spin-flip Metropolis implementation that acts directly on the HUBO cost function of Eq.~\eqref{eq:hubo}, with the cubic terms included in the local fields, so that no quadratization or auxiliary variables are needed~\cite{suzuki_simulated_2025}. One sweep visits the $m$ spins in sequence, and the inverse temperature increases geometrically over the sweeps. The largest possible energy increase of a single flip is accepted with probability $1/2$ and the smallest one is accepted with probability $10^{-2}$. Each read starts from uniformly random spins and the best of all reads is kept. For the heavy-hex problems, with and without random cubic terms, $10^4$ reads and $10^4$ sweeps are used, and $10^3$ reads and $10^3$ sweeps for the Gset instances.
    \item Pauli Correlation Encoding (\textbf{PCE})~\cite{sciorilli_towards_2025} is a variational quantum algorithm that encodes the problem's spin variables $z_i$ into a distinct $k$-body Pauli string $\Pi_i$ on $n$ qubits. The value of each variable is read out as the sign of its expectation value, $z_i=\mathrm{sgn}\langle\Pi_i\rangle$, so that $m$ variables require a number of qubits that scales as $m^{1/k}$, thus requiring fewer qubits than variables. To find the optimal solution, the parameters of a variational quantum circuit are optimized with a classical solver to minimize the problem loss function where the variables are replaced by the expectation values of the Pauli strings. The solution obtained by the variational routine is further refined by local search. PCE is used as a reference for the Gset instances, and its reported results are taken from~\cite{sciorilli_towards_2025}.
    \item The \textbf{greedy} baseline is a multistart local search. Starting from uniformly random spins, it repeatedly flips the spin whose value change gives the largest decrease of the energy of the HUBO Hamiltonian, until no single flip lowers the energy. This is repeated from $10^5$ independent random spin strings, and the lowest energy is kept as the best solution.

\end{enumerate}

\section*{\texorpdfstring{Data availability}{Data availability}}

The datasets used and/or analyzed during the current study are available from the corresponding author on reasonable request.

\section*{Acknowledgements}
We acknowledge the use of IBM Quantum Credits for this work. The views expressed are those of the authors, and do not reflect the official policy or position of IBM or the IBM Quantum team. This research has been supported by ARQADE (CER-20251019), funded by the CERVERA Research Programme of CDTI (Centre for Technological Development and Innovation).

\section*{\texorpdfstring{AI disclosure statement}{AI disclosure statement}}
The ideation and algorithm design was done by the authors. Claude Code (Opus 5.5) was used to facilitate code writing and manuscript polishing, as well as to improve the style of the figures. In addition, Claude Code (Opus 5.5) assisted in editing the appendix. The authors assume full responsibility for the content of this article.

\newpage

\printbibliography 

\newpage

\onecolumn
\appendix
\renewcommand{\thesubsection}{\thesection.\arabic{subsection}} 

\section{Number of QAOA layers}
\label{app:layers}

This appendix collects the tests used to set an adequate number of layers $p$ of the QAOA sampler for each problem and number of qubits in the hardware runs of Section~\ref{sec:hardware_results}. Each test is a single QSSD iteration on \texttt{ibm\_boston} for different values of $p$, with $10^4$ shots and the same transpilation as in Section~\ref{sec:circuit_optimization}. The ratio of the mean sampled energy and the ratio of the lowest energy found are reported, both normalized following Eq.~\eqref{eq:r_scale}.

\begin{figure}[ht]
\centering
\includegraphics[width=\textwidth]{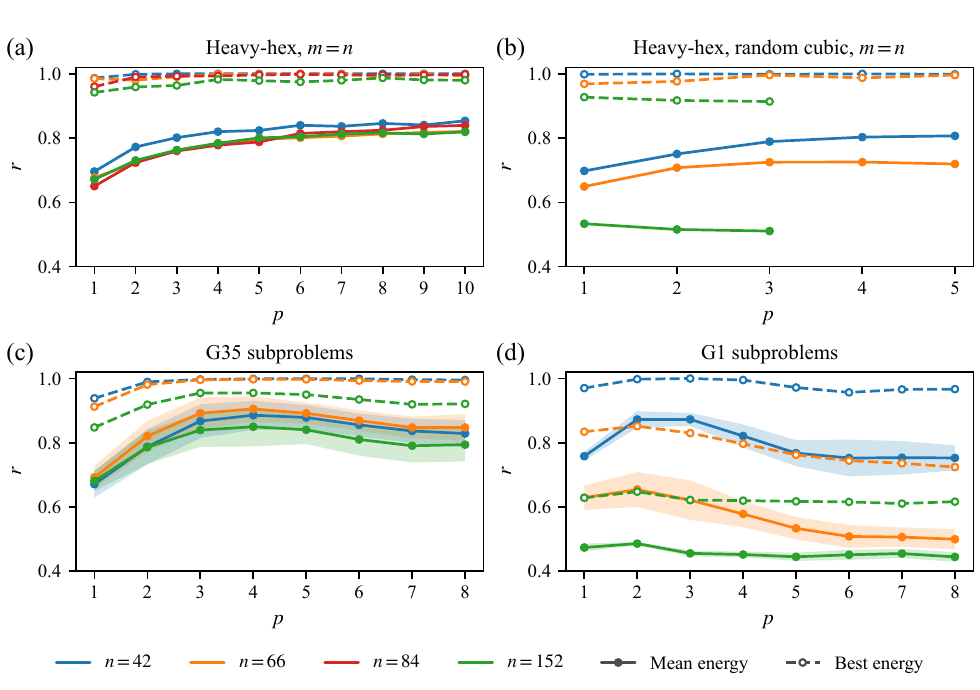}
\caption{Approximation ratio $r$ of the mean (solid) and best (dashed) sampled energy of a single QSSD iteration on \texttt{ibm\_boston} as a function of the number of QAOA layers $p$, for different numbers of qubits $n$. (a) Heavy-hex HUBO (problem class i)) and (b) heavy-hex HUBO with random cubic terms (problem class iii)), with $m=n$, so that the circuit acts on the complete instance. Each point corresponds to one job and the best energy is taken after the bitflip post-process. (c) G35 and (d) G1 subproblems of $n$ randomly selected variables, with the solid line representing the mean over 10 subsets, with the shaded area showing $\pm1$ standard deviation, and the dashed line the best energy without post-process. Uniform sampling gives $r\approx0.5$ in all panels. For the random-cubic instance with $n=152$, only $p\le3$ was run, as even for $p=1$ the results were mostly noise.}
\label{fig:app_layers}
\end{figure}

For the two heavy-hex problems, classes i) and iii) of Section~\ref{sec:problem_classes}, the circuit acts on a complete instance, with $m=n$, with one job per value of $p$. The best energy is taken after the bitflip post-process of the $S_p=100$ lowest-energy samples. For the Gset instances G1 and G35, the test is carried out on 10 random subsets of $n\in\{42,66,152\}$ variables, with the remaining variables frozen to their initial value and the same subsets used for every $p=1,\dots,8$. In this case, $r$ is computed with the bounds of each subproblem obtained with HiGHS, and the best energy is the lowest sampled energy, without post-processing. In all cases, uniform sampling gives $r\approx0.5$. The results are shown in Fig.~\ref{fig:app_layers}.

For the heavy-hex HUBOs without random cubic terms, Fig.~\ref{fig:app_layers}a, the mean ratio increases with $p$ up to $p=10$ for all sizes, with little gain beyond $p=6$. The number of layers is therefore set by the best energy, which is the quantity that QSSD propagates. For $n=42$ the best sample is the ground state from $p=4$, and $p=5$ is used. For $n=66$, the best energy after the post-process reaches the ground state at $p=6$, and the heavy-hex ($m=200$) runs use $p=5$ and $p=7$. For $n=84$ and $n=152$, the best energy is highest at $p=9$ and $p=8$, respectively, which are the values used for heavy-hex ($m=200$) and heavy-hex ($m=10000$). The number of layers needed to obtain better solutions thus increases with $n$, which is expected. Since the best energy fluctuates from one $p$ to the next and each point is a single job, the selected values indicate an adequate reference rather than an optimum.

With random cubic terms (Fig.~\ref{fig:app_layers}b), for $n=42$, the mean ratio increases up to $p=5$ and the post-processed best reaches the ground state at $p=4$. For $n=66$, the mean ratio saturates from $p=3$ and the best energy after the post-process is highest at $p=3$, which is the value used in Section~\ref{sec:hardware_results} and in Appendix~\ref{app:time_cost}. Using larger $p$ does not improve the results. For $n=152$, the mean ratio is already close to uniform sampling at $p=1$ and decreases with $p$, so this size was not used for this problem class.

For G35, Fig.~\ref{fig:app_layers}c, the mean ratio of the subproblems with $n=42$, $66$ and $152$ is highest at $p=4$ and decreases for larger $p$, and the best energies are very close or equal to the subproblem optimum at $p=4$ for the smaller sizes. A value of $p=4$ was therefore used for all sizes of G35 and for the run on \texttt{ibm\_miami} with $n=42$, on which no layer test was carried out due to a lack of resources.

G1, Fig.~\ref{fig:app_layers}d, is a dense graph, and its subproblems have higher depth than the sparser G35 (also seen in Appendix~\ref{app:transpilation}). For $n=42$, the mean ratio is highest at $p=2$--$3$ and again the best sample is close or equal to the subproblem optimum for both. Thus, the run of Section~\ref{sec:hardware_results} uses $p=2$. For $n=66$, the mean ratio is already low at $p=2$ and falls to the uniform value for $p\geq6$, and for $n=152$ it is below the uniform value for every $p$. The hardware run of G1 in Section~\ref{sec:hardware_results} therefore uses $n=42$.

Overall, for problems whose subproblems are compatible with the device topology, the noise is not limiting up to $p=10$ for the sizes tested, and $p$ increases with $n$. For problems whose subproblems require SWAP gates, the best $p$ is between 2 and 4, and for the largest $n$ the samples are close to, or worse than, uniform sampling for any $p$. These tests use a single iteration from a uniform or all-ones reference, so the selected values are a guide for the hardware runs, not an optimum for every iteration.

\section{QAOA cost landscape of the QSSD subproblems}
\label{app:landscape}

\begin{figure}[t]
\centering
\includegraphics[width=0.95\textwidth]{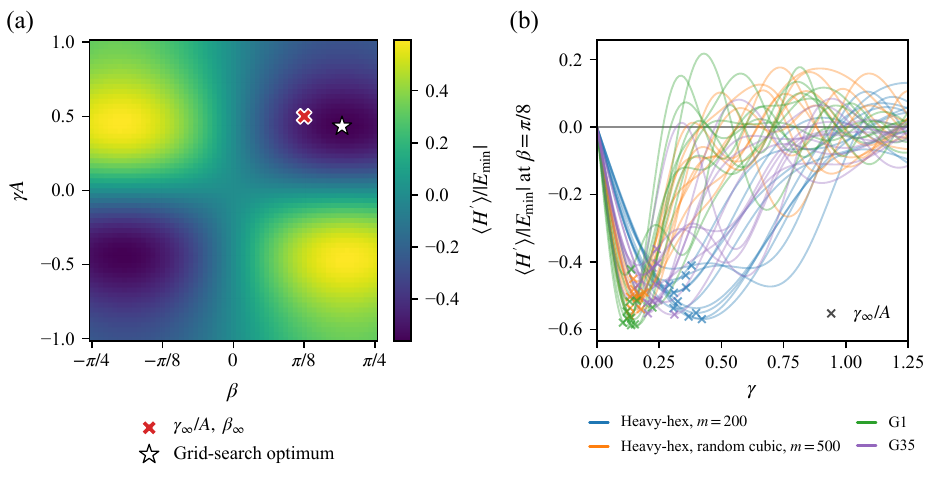}
\caption{Ideal $p=1$ QAOA landscape of QSSD subproblems ($n=20$) of heavy-hex ($m=200$), of the heavy-hex instance with random cubic terms ($m=500$), and of the Gset instances G1 and G35. The mean energy is normalized as $\langle H'\rangle/\lvert E_\mathrm{min}\rvert$, which is 0 for uniform sampling and $-1$ for the ground state. (a) Landscape averaged over one subproblem of each problem, as a function of $\beta$ and of $\gamma$ rescaled by the coefficient magnitude $A$ of each subproblem ($61\times61$ grid). The red cross marks the angles used by QSSD, $(\gamma A,\beta)=(\gamma_\infty,\beta_\infty)=(0.5,\pi/8)$, and the white star the grid-search optimum with $\gamma>0$. The landscape is symmetric under $(\gamma,\beta)\to(-\gamma,-\beta)$. (b) Mean energy at fixed $\beta=\beta_\infty$ as a function of $\gamma$ for the 10 subproblems of each problem. The crosses mark the angle used by QSSD, $\gamma=\gamma_\infty/A$.}
\label{fig:app_landscape}
\end{figure}

This appendix checks how close the fixed angles used by QSSD are to the optimum of the ideal $p=1$ QAOA landscape of the subproblems the algorithm works on, to give an intuition of why they work (results showing how the fixed angles also work adequately for greater depths are displayed in Appendix~\ref{app:boltzmann}). For $p=1$, the angle vectors $(\boldsymbol{\gamma}_t,\boldsymbol{\beta}_t)$ of Eq.~\eqref{eq:parameters} reduce to two scalars, which we write $\gamma=\gamma_\infty/A$ and $\beta=\beta_\infty$, dropping the iteration index $t$ as for the reduced Hamiltonian $H'$. Here $\gamma_\infty=0.5$ and $\beta_\infty=\pi/8$ are the $p=1$ values of $(\boldsymbol{\gamma}_\infty,\boldsymbol{\beta}_\infty)$, and $A=A_t$ is the root-mean-square coefficient of the reduced Hamiltonian defined in Eq.~\eqref{eq:A}. The instances considered are heavy-hex ($m=200$), the heavy-hex instance with random cubic terms and $m=500$, and the Gset instances. For each of them, 10 subproblems of $n=20$ variables are generated with the selection step of QSSD from a random variable order, as in the first iteration, and the remaining variables are frozen to a random reference string. For each reduced Hamiltonian $H'$, the expectation value $\langle H'\rangle(\gamma,\beta)$ is computed exactly by state vector emulation, without shot noise or hardware noise, and agrees with the circuit sent to the device to within $7\cdot10^{-12}$. The constant offset is left out, and the expectation value is normalized by the magnitude of the ground-state energy of $H'$, $\varepsilon=\langle H'\rangle/\lvert E_\mathrm{min}\rvert$. Uniform sampling thus gives $\varepsilon=0$ and a sampler that returns only the ground state gives $\varepsilon=-1$.

Figure~\ref{fig:app_landscape}a shows the landscape as a function of $\beta$ and of the rescaled angle $\gamma A$, averaged over one subproblem of each problem. It has a single broad minimum that contains the QSSD angles. The grid-search optimum, at $\gamma A=0.43$ and $\beta=0.60$, gives $\varepsilon=-0.561$, against $-0.477$ at the QSSD angles, meaning the QSSD angles achieve 85\% of the optimal energy decrease.

Fixing $\beta=\beta_\infty$, Fig.~\ref{fig:app_landscape}b, the QSSD rescaled angles achieve on average 98\% of the energy decrease of the best $\gamma$, and at least 87\% for every subproblem (Table~\ref{tab:app_landscape}). Optimizing $\beta$ as well, the QSSD angles achieve on average 90\% of the optimal energy decrease for heavy-hex ($m=200$) and between 72\% and 78\% for the other three problems. The rescaling by $A$ therefore removes most of the variation of the optimal $\gamma$ between subproblems, and the remaining loss with respect to the optimal $p=1$ angles comes mainly from keeping $\beta$ fixed.

\begin{table}[t]
\centering
\small
\renewcommand{\arraystretch}{1.2}
\begin{tabular}{l ccc}
\toprule
\textbf{Problem} & $\gamma_\infty/A$, $\beta_\infty$ (QSSD) & $\gamma_\mathrm{opt}$, $\beta_\infty$ & $\gamma_\mathrm{opt}$, $\beta_\mathrm{opt}$ (grid search) \\
\midrule
Heavy-hex ($m=200$)            & $-0.502 \pm 0.054$ & $-0.514 \pm 0.043$ & $-0.555 \pm 0.043$ \\
Heavy-hex, random cubic ($m=500$)  & $-0.508 \pm 0.032$ & $-0.510 \pm 0.031$ & $-0.655 \pm 0.074$ \\
G1                     & $-0.542 \pm 0.050$ & $-0.547 \pm 0.048$ & $-0.750 \pm 0.067$ \\
G35                    & $-0.478 \pm 0.066$ & $-0.494 \pm 0.050$ & $-0.651 \pm 0.072$ \\
\bottomrule
\end{tabular}
\caption{Mean energy at the QSSD angles and at the optimal $p=1$ angles, averaged over the 10 subproblems of each problem, with the best $\gamma$ taken at fixed $\beta=\beta_\infty$ or with $\gamma$ and $\beta$ optimized together. The values are the mean $\pm$ the standard deviation over the 10 subproblems.}
\label{tab:app_landscape}
\end{table}

\section{Low-energy enhancement of the QAOA sampler}
\label{app:boltzmann}

\begin{figure}[t]
\centering
\includegraphics[width=\textwidth]{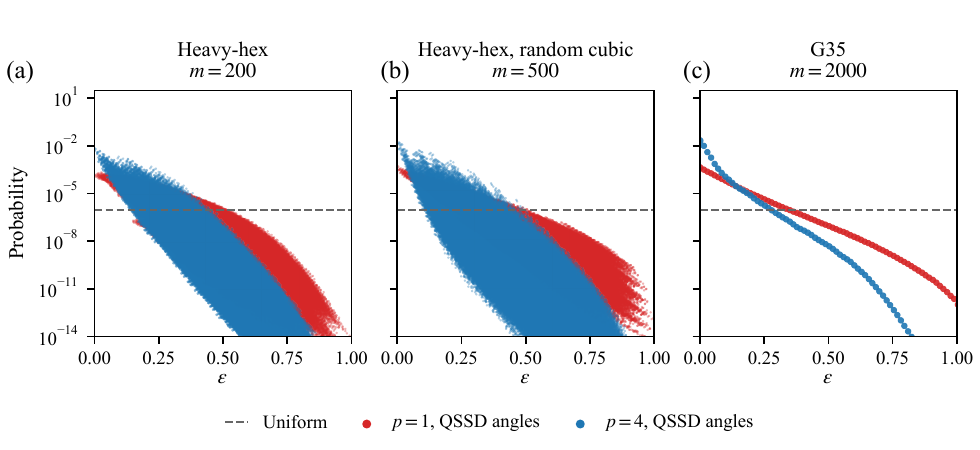}
\caption{Probability of sampling one configuration as a function of its normalized energy $\epsilon$, for the QAOA state with the QSSD angles and $p=1$ (red) and $p=4$ (blue), computed exactly for one subproblem with $n=20$ of (a) heavy-hex ($m=200$), (b) the heavy-hex instance with random cubic terms ($m=500$) and (c) G35. Each point is one energy level, with the probability of the level divided by its number of configurations; G35 has only a few, highly degenerate, levels. The dashed line is uniform sampling and the ground state is at $\epsilon=0$.}
\label{fig:app_boltzmann}
\end{figure}

This appendix shows that the QAOA sampler of QSSD shifts the sampled distribution towards low energies for tested HUBO problems and with the fixed, suboptimal angles of Eq.~\eqref{eq:parameters}, and that adding layers with these same angles increases the probability of sampling low-energy configurations further. One subproblem of $n=20$ variables of Appendix~\ref{app:landscape} is taken for each class of problem, heavy-hex ($m=200$), the heavy-hex instance with random cubic terms ($m=500$) and G35, and the exact output distribution of the QAOA circuit with the QSSD angles is computed for $p=1$ and $p=4$. Each configuration is characterized by its normalized energy $\epsilon=(E-E_\mathrm{min})/(E_\mathrm{max}-E_\mathrm{min})\in[0,1]$, where $E_\mathrm{min}$ and $E_\mathrm{max}$ are the lowest and highest energies of the subproblem, so that $\epsilon=0$ is the ground state and $\epsilon=1$ the highest-energy configuration.

For the three problems, the probability of sampling a configuration decreases with its energy, so that configurations of low energy are sampled more often than with uniform sampling and those of high energy are suppressed by several orders of magnitude (Fig.~\ref{fig:app_boltzmann}). The behavior is the same for the quadratic problem (G35) and for the two HUBO problems. With $p=1$, the probability of sampling a configuration with $\epsilon\le0.1$ grows from about $0.1\%$ for uniform sampling to 10--14\%, although the $p=1$ QSSD angles are not optimal for these subproblems (Appendix~\ref{app:landscape}). With $p=4$ and the same rule for the angles, which are not optimized for each subproblem either, the decrease with the energy becomes steeper, the configurations near the ground state become more likely than with $p=1$, and this probability reaches 45--56\%, and the probability of the ground state is about $5000$ to $22000$ times that of uniform sampling, against about 200 to 500 times for $p=1$. Adding layers therefore keeps improving the sampler even with suboptimal angles.

\section{Runtime of the classical baselines}
\label{app:time_cost}

Table~\ref{tab:app_baseline_times} lists the wall-clock time each classical baseline took on each problem instance mentioned throughout the results reported in Section~\ref{sec:hardware_results}, with the settings of Section~\ref{sec:classical_quantum_baseline}. The runs were not designed as a strict timing benchmark. The times given are therefore only a reference for the cost of each baseline. The CPU used is an Intel Core i7-13650HX. Looking at the measured times, the cost of each algorithm depends more on the structure of the instance than just on the number of variables. HiGHS is the fastest method on the heavy-hex instances without random cubic terms, where it proves optimality, but becomes the slowest by several orders of magnitude on the Gset instances and on the instances with random cubic terms. The time of SA and of the greedy baseline is set by their budget of reads, sweeps or restarts.

\begin{table}[ht]
\centering
\small
\renewcommand{\arraystretch}{1.2}
\setlength{\tabcolsep}{6pt}
\begin{tabular}{l r rrr}
\toprule
\textbf{Instance} & $m$ & \textbf{HiGHS} & \textbf{SA} & \textbf{Greedy} \\
\midrule
Heavy-hex                 & 200    & 0.6          & 313           & 5.3 \\
Heavy-hex                 & 10000 & 3310      & $^{\dagger}$3300 & 503 \\
G1                        & 800    & $^{*}$7200 & 37      & 207 \\
G35                       & 2000 & $^{*}$7200 & 52      & 107 \\
Heavy-hex, random cubic   & 200    & 50           & 200           & 2.6 \\
Heavy-hex, random cubic   & 300    & $1.0\cdot10^{5}$ & 316      & 5.2 \\
Heavy-hex, random cubic   & 400    & $^{*}2.6\cdot10^{5}$ & 460 & 8.5 \\
Heavy-hex, random cubic   & 500    & $^{*}2.6\cdot10^{5}$ & 643 & 13.9 \\
\bottomrule
\end{tabular}
\caption{Time in seconds of a single run of each classical baseline. HiGHS, SA and greedy run on one thread, except SA on heavy-hex ($m=10000$), marked $\dagger$, which runs its reads in parallel on 8 threads. The runs marked with $^{*}$ stopped at the time limit without proving optimality.}
\label{tab:app_baseline_times}
\end{table}

\section{Quantum hardware}
\label{app:hardware}

All the hardware experiments were run on IBM superconducting processors through Qiskit IBM Runtime, whose characteristics are summarized in Table~\ref{tab:app_hardware}. \texttt{ibm\_boston} is a Heron processor with the heavy-hex lattice, in which every qubit is coupled to at most three neighbors. \texttt{ibm\_miami} is a Nighthawk processor with a square lattice, in which most qubits are coupled to four neighbors. All devices use the native gate set $\{\mathrm{CZ}, R_z, \sqrt{X}, X\}$. The median two-qubit gate errors of these devices are of order $10^{-3}$ and the readout errors of order $10^{-2}$. Since the experiments were carried out over several months, the specific values changed between calibrations. Most of the results were obtained on \texttt{ibm\_boston}, using up to $n=152$ of its 156 qubits. 

Each QSSD iteration executes a single circuit with the sampler primitive, with $10^4$ shots on \texttt{ibm\_boston} and $1024$ shots on \texttt{ibm\_miami}. The circuit is transpiled again at every iteration with the Qiskit preset pass manager at optimization level 3. No error suppression or mitigation was applied in the runs of Section~\ref{sec:hardware_results}. The emulated runs use the noiseless state vector simulator of Qiskit Aer, with the same circuits and no transpilation to a device.

\begin{table}[ht]
\centering
\small
\renewcommand{\arraystretch}{1.2}
\begin{tabular}{lccc}
\toprule
\textbf{Device} & \textbf{Processor} & \textbf{Qubits} & \textbf{Topology} \\
\midrule
\texttt{ibm\_boston}     & Heron r3     & 156 & heavy-hex \\
\texttt{ibm\_miami}      & Nighthawk r1 & 120 & square    \\
\bottomrule
\end{tabular}
\caption{IBM quantum processors used in this work.}
\label{tab:app_hardware}
\end{table}

\section{Complete QSSD runs on quantum hardware}
\label{app:hw_runs}

\begin{figure}[ht]
\centering
\includegraphics[width=\textwidth]{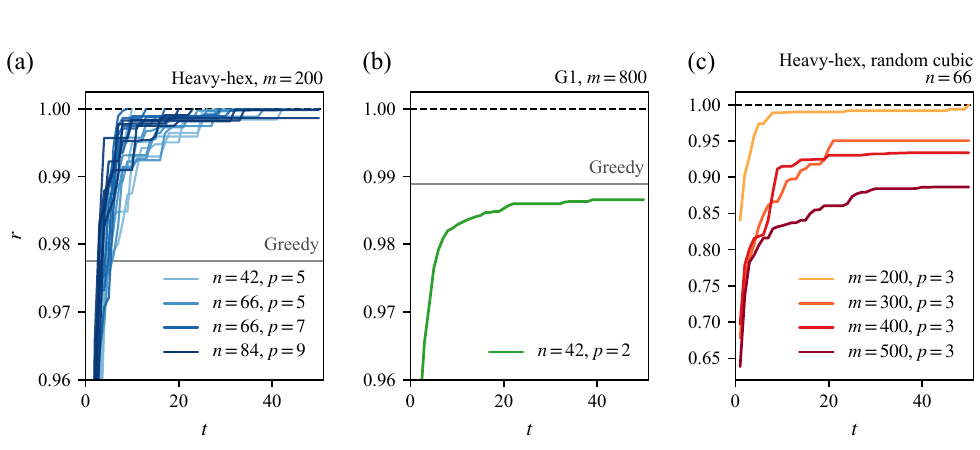}
\caption{Approximation ratio of the reference solution as a function of the iteration $t$ for the QSSD runs on \texttt{ibm\_boston}. (a) Heavy-hex with $m=200$, five runs for each number of qubits $n$ and layers $p$ (Table~\ref{tab:ibm_boston_HUBO_M200}); the dashed line is the optimum and the gray line the greedy baseline. (b) G1, one run with $n=42$ and $p=2$ (Section~\ref{sec:hardware_results}); the dashed line is the best-known cut and the gray line the greedy baseline. (c) Heavy-hex with random cubic terms, one run with $n=66$ and $p=3$ for each $m$ (Table~\ref{tab:hubo_summary}); the dashed line is the best-known energy.}
\label{fig:app_hw_runs}
\end{figure}

Fig.~\ref{fig:app_hw_runs} shows the evolution of the approximation ratio $r$ of the reference solution with the iteration $t$ for the complete QSSD runs on \texttt{ibm\_boston} on the heavy-hex instance with $m=200$ of Table~\ref{tab:ibm_boston_HUBO_M200}, the G1 instance and the heavy-hex with extra random cubic terms in Table~\ref{tab:hubo_summary}. Fig.~\ref{fig:app_hw_runs}a shows the five runs of each configuration on the heavy-hex instance with $m=200$, with $r$ computed with Eq.~\eqref{eq:r_scale}, together with the greedy baseline. Fig.~\ref{fig:app_hw_runs}b shows the single run on G1 with $n=42$ and $p=2$, with $r$ the ratio of the cut to the best-known cut, together with the greedy baseline. Fig.~\ref{fig:app_hw_runs}c includes the runs with $n=66$ and $p=3$ on the heavy-hex instances with random cubic terms and $m\in\{200, 300, 400, 500\}$, with $r$ computed with Eq.~\eqref{eq:r_min}.

\section{Uniform sampling and hardware samples}
\label{app:samplers}

This appendix examines how QSSD behaves when the samples carry no information about the subproblem, since the bitflip post-process acts on all $m$ variables and can improve the reference solution by itself, and compares the samples obtained on hardware with uniform samples on the same subproblems.

QSSD is run with two classical samplers: a uniform sampler, which draws $S=10^4$ random strings on the $n$ selected variables, and an exact sampler, which returns the ground state of each subproblem computed with HiGHS, with a limit of 60\,s per subproblem. The other settings are those of Section~\ref{sec:hardware_results}, with $T=50$ iterations and three seeds per size.

\begin{figure}[ht]
\centering
\includegraphics[width=\textwidth]{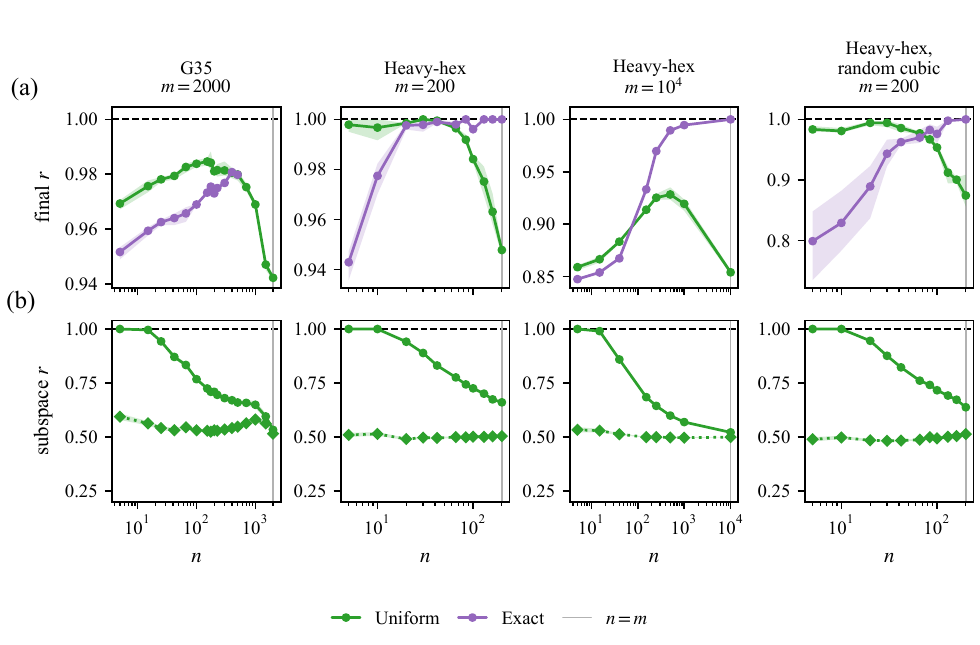}
\caption{QSSD with a uniform (green) and an exact (purple) sampler as a function of the subproblem size $n$. (a) Final approximation ratio after $T$ iterations, mean over three seeds. The exact sampler is shown only where HiGHS proved the optimum of every subproblem. (b) Approximation ratio on the subproblem of the best (solid) and mean (dotted) uniform sample, before the post-process.}
\label{fig:sampler_dip}
\end{figure}

Figure~\ref{fig:sampler_dip}a shows that, with uniform samples, the quality of the final solution first increases with $n$, since each iteration updates more variables, reaches a maximum and then degrades for large $n$. This is not an effect of the fixed number of iterations, but of the samples themselves. Before the post-process, the best uniform sample moves away from the subproblem optimum as $n$ grows (Fig.~\ref{fig:sampler_dip}b), so that for large $n$ the improvement comes from the post-process alone. The exact sampler, instead, keeps improving with $n$ and reaches the best solutions at large $n$, but solving the subproblems exactly becomes very expensive for large and dense subproblems, as shown for $n\ge700$ where HiGHS no longer proves the optimum within the time limit. When the subproblems can be solved exactly and fast, an exact solver is naturally preferable to a heuristic sampler.

\begin{figure}[ht]
\centering
\includegraphics[width=\textwidth]{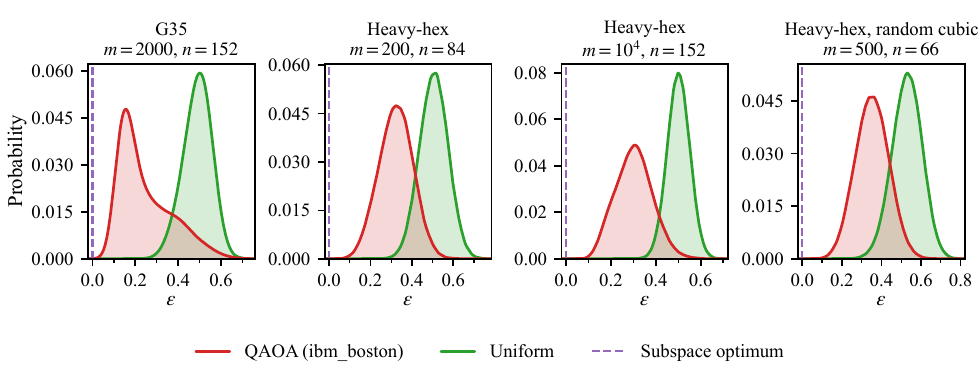}
\caption{Distribution of the normalized energy $\epsilon$ of the samples measured on \texttt{ibm\_boston} (red) and of uniform samples (green) on the same subproblems, for the largest subproblem size run on hardware for each problem class, averaged over the iterations $t\ge1$ of one hardware run (59, 22, 99 and 49 iterations, from left to right). $\epsilon=0$ is the optimum of each subproblem (dashed line) and $\epsilon=1$ its maximum. The curves are kernel density estimates, shown as the probability of an interval of width 0.01 in $\epsilon$.}
\label{fig:hw_distributions}
\end{figure}

To compare the hardware samples with uniform samples on the same subproblems, the subproblem, the reference solution and the $10^4$ samples measured on \texttt{ibm\_boston} are taken at every iteration $t\ge1$ of the runs of Section~\ref{sec:hardware_results}, and $10^4$ uniform samples are drawn on the same variables. Figure~\ref{fig:hw_distributions} shows the distribution of the normalized energy $\epsilon=(E-E_\mathrm{min})/(E_\mathrm{max}-E_\mathrm{min})\in[0,1]$, where $E_\mathrm{min}$ and $E_\mathrm{max}$ are the lowest and highest energies of each subproblem, obtained with HiGHS, so that $\epsilon=0$ is the ground state of the subproblem (as in Appendix~\ref{app:boltzmann}), for the largest hardware size of each problem class. The hardware samples are concentrated at lower energies than the uniform ones for all problems (Fig.~\ref{fig:hw_distributions}), with a mean $\epsilon$ between 0.25 and 0.35, against about 0.5 for uniform samples.

While the subproblems are small, QSSD with uniform sampling is able to reach good solutions, because a set of random strings on a few variables almost always contains a good solution of the subproblem. As the subproblems grow, random strings stop containing good solutions, and the final result degrades. A sampler that solves the subproblems well removes this limit and keeps improving with the subproblem size. The QAOA hardware samples lie between these two cases and are biased towards low energies, so the post-process starts from better strings and improves the reference more often than with random strings, an advantage that grows with the subproblem size.

\section{QAOA circuit transpilation}
\label{app:transpilation}

This appendix details the construction of the QAOA cost layer described in Section~\ref{sec:circuit_optimization} and evaluates its two-qubit cost. The first subsection describes the construction of Ref.~\cite{pelofske_short-depth_2024} for the hardware-native heavy-hex problems, whose cost layer has CNOT depth six. The second subsection compares, on the subproblems that QSSD generates for the heavy-hex and Gset instances, the CZ count and the two-qubit depth after transpilation of the resulting circuits with those of the standard Qiskit \texttt{QAOAAnsatz} circuit, and discusses for which subproblems the depth-six layer applies.

\subsection{Depth-six cost layer for heavy-hex problems}
\label{app:native_depth}

Let $G=(V,\mathcal{E})$ be a subgraph of the heavy-hex lattice. The lattice is bipartite, with one part formed by the vertices of degree two, which we denote $V_2$, and the other by the vertices of degree three, and any subgraph inherits this partition. Following Eq.~(4) of Ref.~\cite{pelofske_short-depth_2024}, a problem is hardware-native on $G$ when it has the form of Eq.~\eqref{eq:hubo} with $J_{ij}\neq0$ only for edges $(i,j)\in\mathcal{E}$ and $K_{ijk}\neq0$ only when $j\in V_2$ and $i$ and $k$ are its two neighbors. Every cubic term is therefore centered on a vertex of $V_2$ and supported on its two edges. 

As shown in Ref.~\cite{pelofske_short-depth_2024}, for such a Hamiltonian $H$ the cost unitary $e^{-i\gamma_l H}$ can be implemented exactly with $2|\mathcal{E}|$ CNOT gates, all acting on edges of $G$, with a CNOT depth of six independently of $|V|$ and of the number of cubic terms.

Since $G$ is bipartite with maximum degree three, its edges can be colored with three colors so that edges of the same color share no vertex. Each edge $(i,j)$ is implemented with CNOTs controlled by $i$ and targeting $j\in V_2$. The cost layer consists of the linear terms, applied as $R_z(2\gamma_l h_i)$ on every qubit, followed by a forward sweep of three layers of CNOTs, one per color, and a backward sweep with the same three layers in the same order, so that every edge contributes one CNOT to each sweep. The quadratic and cubic rotations are applied on the target qubits $j\in V_2$ between these CNOT layers.
In the computational basis, a CNOT multiplies the spin value of its target by that of its control. A rotation $R_z(2\gamma_l J_{ij})$ applied on $j$ while it holds the product $z_iz_j$ therefore produces the phase $e^{-i\gamma_l J_{ij}z_iz_j}$ of the corresponding term, and similarly for the cubic terms. If $j$ has a single neighbor $i$, $R_z(2\gamma_l J_{ij})$ is applied right after the forward CNOT on $(i,j)$. If $j$ has two neighbors $i$ and $k$, labeled so that the color of $(i,j)$ comes first, the forward CNOTs on $(i,j)$ and $(k,j)$ and the backward CNOTs on $(i,j)$ and $(k,j)$ change the value held by $j$ as
\begin{equation*}
    z_iz_j \;\to\; z_iz_jz_k \;\to\; z_jz_k \;\to\; z_j ,
\end{equation*}
and the rotations $R_z(2\gamma_l J_{ij})$, $R_z(2\gamma_l K_{ijk})$ and $R_z(2\gamma_l J_{jk})$ are applied after the first three of these CNOTs, respectively.

Edges of the same color share no vertex, so each sweep has depth three and the cost layer contains $2|\mathcal{E}|$ CNOTs at depth six. All controls lie outside $V_2$ and all targets in $V_2$, so the CNOTs commute, and since each appears twice the CNOT network returns every basis state to itself. The circuit therefore equals $e^{-i\gamma_l H}$ exactly, and each cubic term costs a single $R_z$ gate and no additional CNOT.

\subsection{Transpiled cost against the standard Qiskit \texttt{QAOAAnsatz} circuit on QSSD subproblems}
\label{app:transpilation_subproblems}

The reduced Hamiltonians $H'_t$ of Eq.~\eqref{eq:H_eff} that QSSD sends to the device, written $H'$ in this appendix, are in general not hardware-native problems on a subgraph of the heavy-hex lattice. For each $H'$, the QSSD circuit is built together with the circuit of the Qiskit \texttt{QAOAAnsatz} class~\cite{javadi-abhari_quantum_2024}, in which each term of $H'$ is synthesized as an independent Pauli evolution. Both implement the same QAOA state, compile both with the same preset pass manager at optimization level 3, and compare the number of CZ gates and the two-qubit depth.

The subproblems are the $n$-variable subproblems selected in the first iteration of QSSD, for $n\in\{20, 42, 66, 100, 152\}$ and five random seeds. The instances considered are the heavy-hex ($m=200$) instance, the heavy-hex instance with random cubic terms and $m=200$, and the Gset instances G1 and G35. The targets are the coupling maps of \texttt{ibm\_boston} (heavy-hex) and, for the Gset instances, \texttt{ibm\_miami} (square lattice), with native gate set $\{\mathrm{CZ}, R_z, \sqrt{X}, X\}$ and no calibration data. All circuits have $p=1$. For $n=8$, exact state vector emulation gives the same output distribution for both circuits, with a total variation distance below $1.1\cdot10^{-15}$.

\begin{figure}[ht]
\centering
\includegraphics[width=\textwidth]{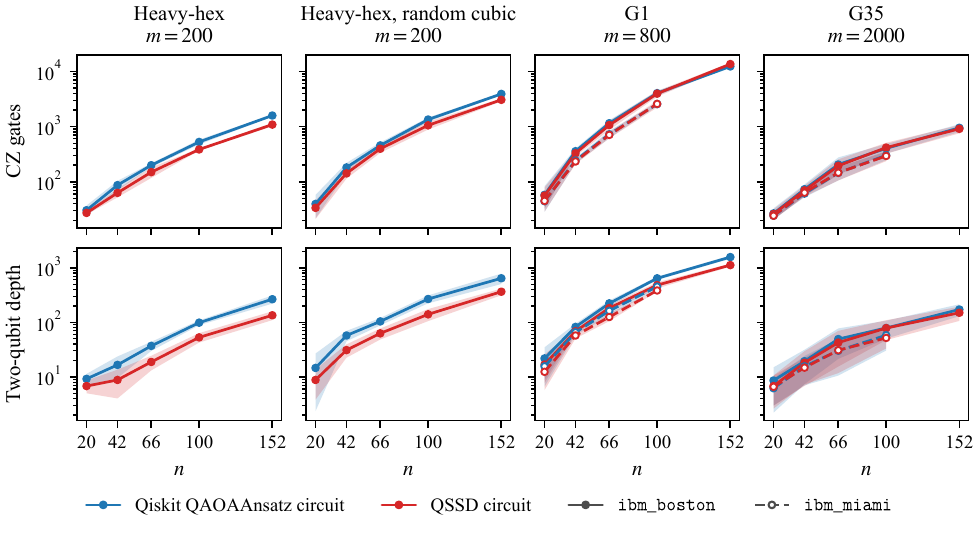}
\caption{CZ count (top) and two-qubit depth (bottom) after transpilation at optimization level 3, for the QSSD circuit construction and for the Qiskit \texttt{QAOAAnsatz} circuit applied to the same reduced Hamiltonians, as a function of the subproblem size $n$ ($p=1$). Solid lines and filled markers correspond to the heavy-hex \texttt{ibm\_boston}, and dashed lines and open markers to the Gset instances transpiled to the square-lattice \texttt{ibm\_miami}. Markers show the mean over five subproblems selected by QSSD, and error bars one standard deviation.}

\label{fig:app_transpilation}
\end{figure}

\begin{table}[ht]
\centering
\small
\renewcommand{\arraystretch}{1.2}
\setlength{\tabcolsep}{4pt}
\begin{tabular}{ll ccccc}
\toprule
 & & \multicolumn{5}{c}{$n$} \\
\cmidrule(lr){3-7}
\textbf{Problem} & \textbf{Device} & 20 & 42 & 66 & 100 & 152 \\
\midrule
Heavy-hex ($m=200$)            & \texttt{ibm\_boston} & 1.12 / 1.48 & 1.38 / 1.98 & 1.40 / 2.22 & 1.37 / 1.92 & 1.46 / 2.03 \\
Heavy-hex, random cubic ($m=200$)  & \texttt{ibm\_boston} & 1.16 / 1.61 & 1.28 / 1.89 & 1.15 / 1.70 & 1.29 / 1.99 & 1.28 / 1.80 \\
G1                     & \texttt{ibm\_boston} & 1.02 / 1.29 & 1.10 / 1.21 & 1.09 / 1.23 & 1.02 / 1.34 & 0.91 / 1.40 \\
G35                    & \texttt{ibm\_boston} & 1.02 / 1.16 & 1.05 / 1.08 & 1.04 / 1.30 & 0.98 / 1.08 & 1.04 / 1.18 \\
G1                     & \texttt{ibm\_miami}  & 0.99 / 1.33 & 1.04 / 1.16 & 1.02 / 1.30 & 1.00 / 1.18 & -- \\
G35                    & \texttt{ibm\_miami}  & 1.00 / 0.97 & 0.96 / 1.04 & 1.00 / 0.97 & 1.03 / 1.09 & -- \\
\bottomrule
\end{tabular}
\caption{Reduction factors of the CZ count / two-qubit depth obtained with the QSSD circuit construction, defined as the \texttt{QAOAAnsatz} value divided by the QSSD value for the same subproblem and averaged over five subproblems. Values above one favor the QSSD construction.}
\label{tab:app_transpilation}
\end{table}

The largest gains are obtained for the heavy-hex problems (Fig.~\ref{fig:app_transpilation}, Table~\ref{tab:app_transpilation}). For heavy-hex ($m=200$) the QSSD construction reduces the CZ count by a factor of 1.1 to 1.5 and the two-qubit depth by a factor of 1.5 to 2.2. With random cubic terms, the CZ count is reduced by a factor of 1.2 to 1.3 and the depth by a factor of 1.6 to 2.0. The saving in CZ gates comes from the cubic terms supported on the topology, which share the CNOTs of their quadratic terms, so that the QSSD circuit has fewer two-qubit gates before transpilation.

The Gset instances have no cubic terms and their subproblems do not follow the device topology. After transpilation, the CZ counts of the two constructions agree within about 10\% on both devices. The ordering of the rotations by edge coloring still reduces the two-qubit depth, by a factor of 1.2 to 1.4 for G1 on both devices and of 1.1 to 1.3 for G35 on \texttt{ibm\_boston}, while for G35 on \texttt{ibm\_miami} the two constructions are equivalent. For non-native problems, the construction offers no systematic advantage in gate count over the default compilation, as expected.

The depth-six cost layer requires the interaction graph of $H'$ to be bipartite, with one side of its 2-coloring containing only vertices of degree at most two, and every cubic term to be centered on that side. This holds only for the smallest subproblems, such as all five heavy-hex ($m=200$) subproblems with $n=20$, and fails for most larger ones. The selected variables typically form several disconnected fragments of the lattice. The construction itself does not require connectedness, but the implementation chooses the 2-coloring of each fragment independently and then requires a single side of the bipartition with only vertices of degree at most two, which may fail even if every fragment admits a valid assignment. In addition, the cubic terms with one frozen variable add effective quadratic couplings between variables that are not lattice neighbors, which can make the interaction graph non-bipartite. When the nested construction does not apply, the cubic terms are emitted as separate CNOT blocks, and the depth reduction in Table~\ref{tab:app_transpilation} comes from the edge-coloring order and from the cubic terms that share the CNOTs of their quadratic terms.

\end{document}